\documentclass[aps,prb,twocolumn,groupedaddress,showpacs]{revtex4}
\usepackage{graphicx,amsmath}
\usepackage{amssymb}
\usepackage{sidecap}
\usepackage{hyperref}

\begin{document}
\title{Interaction of W(CO)$_6$ with SiO$_2$ Surfaces -- 
A Density Functional Study}

\author{Kaliappan Muthukumar}
\affiliation{Institut f\"ur Theoretische Physik, Goethe-Universit\"at Frankfurt, Max-von-Laue-Stra{\ss}e 1, 60438 Frankfurt am Main, Germany}
\author{Ingo Opahle}
\affiliation{Institut f\"ur Theoretische Physik, Goethe-Universit\"at Frankfurt, Max-von-Laue-Stra{\ss}e 1, 60438 Frankfurt am Main, Germany}
\author{Juan Shen}
\affiliation{Institut f\"ur Theoretische Physik, Goethe-Universit\"at Frankfurt, Max-von-Laue-Stra{\ss}e 1, 60438 Frankfurt am Main, Germany}
\author{Harald O. Jeschke}
\affiliation{Institut f\"ur Theoretische Physik, Goethe-Universit\"at Frankfurt, Max-von-Laue-Stra{\ss}e 1, 60438 Frankfurt am Main, Germany}
\author{Roser Valent\'\i}
\affiliation{Institut f\"ur Theoretische Physik, Goethe-Universit\"at Frankfurt, Max-von-Laue-Stra{\ss}e 1, 60438 Frankfurt am Main, Germany}

\date{\today}

\newcommand{\wc}{W(CO)$_6$}
\newcommand{\mc}{M(CO)$_6$}
\newcommand{\bc}{${\beta}${-}cristobalite}
\newcommand{\sio}{SiO$_2$}
\newcommand{\FOH}{FOH - SiO$_2$ }
\newcommand{\POH}{POH - SiO$_2$ }

\begin{abstract}

The interaction of tungsten hexacarbonyl {\wc} precursor molecules with {\sio} substrates is investigated by means of density functional
theory calculations with and without inclusion of long range van der Waals interactions. We consider two different surface
models, a fully hydroxylated and a partially hydroxylated {\sio} surface, corresponding to substrates under different experimental
conditions.  For the fully hydroxylated surface we observe only a weak interaction between the precursor molecule and the substrate 
with physisorption of {\wc}. Inclusion of van der Waals corrections results in a stabilization of the molecules on this surface, 
but does not lead to significant changes in the chemical bonding. In contrast, we find a spontaneous dissociation of the precursor
molecule on the partially hydroxylated {\sio} surface where chemisorption of a W(CO)$_5$ fragment is observed upon removal of one
of the CO ligands from the precursor molecule.
  Irrespective of the hydroxylation, the precursor molecule prefers binding of 
more than one of its CO ligands. In the light of these results, implications for the initial growth stage of tungsten nano-deposits 
on {\sio} in an electron beam induced deposition process are discussed.
\end{abstract}


\maketitle

\section{Introduction\label{Introduction}}
The growth of patterned transition metal nano-deposits with tunable
electronic properties has been in the focus of recent research towards
nanoscale device applications~\cite{Wnuk2011,Utke2008}. One of the
promising routes to achieve this goal is the preparation of
nano-deposits from gas phase precursor molecules like tungsten
hexacarbonyl ({\wc}) or cyclopentadienyltrimethylplatinum(IV) by means
of focused ion or electron beam induced deposition (EBID) or laser
induced deposition techniques~\cite{Arumainayagam2010,Seuret2003,Utke2008}. 
While experimentally substantial progress has been made in the fabrication
of transition metal nano-deposits with properties reaching from insulating to metallic 
or even superconducting behavior~\cite{Guillamon2008}, a detailed understanding of the growth
processes on a microscopic level is still missing. This is due to a multitude of complicated 
processes involved in the growth of nano-deposits from the gas phase -- like the adsorption or
chemisorption of precursor molecules, heat transfer and the interaction with primary and secondary electrons in the EBID process
as well as fragmentation and recombination of fragments -- which do not allow a one step simulation of the growth process within
reasonable time.
 
{\wc} is one of the most prominent precursor materials used in the EBID process and has been deposited on various substrates like
{\sio}~\cite{Koops1988,Okuyama1980}, W(110)~\cite{Flitsch1991}, and Ni(100)~\cite{Zaera1992} by using a variety of techniques.  The
chemical and physical properties of {\wc} as a molecule are well studied both experimentally and
theoretically~\cite{Brenner1979,Linden1980,Kuzusaka1980,Zaera1992,Ehlers1994,Kurhinen1994,Suvanto1999}.
However, to the best of our knowledge, there haven't been any reports of electronic structure calculations 
that treat the interaction of {\wc} molecules with the {\sio}
substrate. Therefore, in this work, we
focus on constructing a reasonable model for the \sio\
substrate and study the interaction 
of a single {\wc} molecule with the {\sio} substrate
as one important step towards the simulation of a realistic growth process in EBID.

For the simulation of the {\wc} interaction with the {\sio} substrate we employ two different surface models, 
a fully hydroxylated (FOH) and a
partially (POH) hydroxylated {\sio} surface. The \FOH 
represents a realistic model for a substrate prepared under non-vacuum
conditions in the absence of irradiation, while the \POH corresponds to a substrate under the influence of
an electron beam, where a partial removal of OH groups from the
surface is expected due to the interaction with the beam, similar to
other surfaces such as {TiO$_2$} and ${\gamma}$-{Al$_2$O$_3$}~\cite{zhang,Wang2004}. 
The interaction between the precursor molecule and the substrate is then studied
within density functional theory (DFT) using the generalized gradient
approximation (GGA) with and without van der Waals (vdW) interactions.
The impact of vdW interactions, which were found to be important for the
interaction of organic molecules on metal surfaces in recent 
studies~\cite{Atodiresei2009,Atodiresei2010}, is discussed for the
present system.

\section{Computational Details}\label{Formalism}

DFT calculations for bulk {\sio}, isolated {\sio} substrates and
{\wc} precursor molecules, as well as the combined
substrate/precursor molecule system were performed using the projector
augmented wave (PAW) method~\cite{Bloechl1994,Kresse1999} as
implemented in the Vienna Ab-initio Simulation Package
(VASP-5.2.11)~\cite{Kresse1996,Kresse1993,Kresse1994}. GGA in the parametrization of
Perdew, Burke and Ernzerhof~\cite{Perdew1996,Perdew1997} was
used as approximation for the exchange and correlation functional.  In
addition, dispersion corrections~\cite{Grimme2006,Wu2001} were used 
that simulate the long range vdW interactions, which are expected to be important for the adsorption of
the {\wc} molecules on the \FOH  surface.

\begin{figure}  
\begin{center}  
\includegraphics[width=0.50\textwidth]{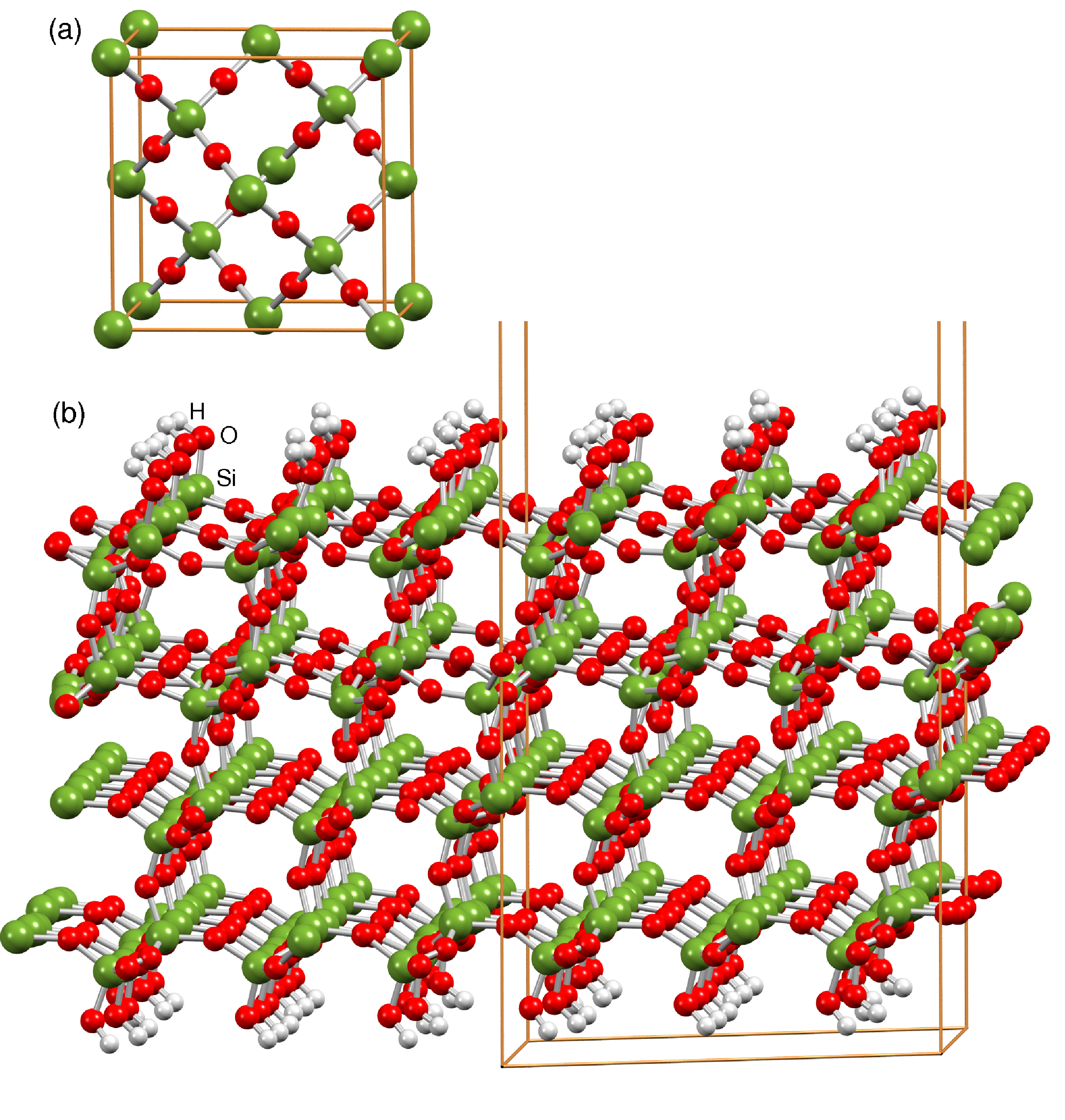}  
\caption{\label{fig:substrate} (Color online)
  (a) Structure of bulk
  $\beta$-cristobalite {\sio} and (b) side view of the slab geometry
  with a (111) surface used for this study. The corresponding unit cell
is indicated.}
\end{center}   
\end{figure}

For bulk {\sio}, the ideal $\beta$-cristobalite structure~\cite{Wyckoff1963} with cubic Fd\={3}m symmetry 
(Fig.~\ref{fig:substrate}~(a)) and the experimental lattice constant $a = 7.16$ \AA\ was used.  
To simulate the substrate-precursor interaction, slabs consisting of 2 to 5 layers
of {\sio} with a hexagonal {\sio} (111) surface were generated from the experimental bulk structure.
The minimal slab geometry to simulate the hexagonal {\sio} (111) surface has in plane lattice parameters corresponding to the primitive 
lattice constants $a_p=a/\sqrt{2}=5.06$ \AA\ of the face centered cubic cell.
This minimal geometry does however not allow for a sufficient separation of adjacent images when the molecule is placed on the surface.
Thus we used a  $3\times 3$ supercell geometry in the plane 
(in terms of primitive lattice constants) for all slab calculations.

For the \FOH substrate, all surface silicon atoms (top and bottom) were terminated with OH groups
(Fig.~\ref{fig:substrate}~(b)).   
In the case of the \POH surface some of the OH groups
were removed from this slab. We considered three cases depending on the precursor
molecule orientation with OH-vacancy concentrations of 11$\%$, 22$\%$ and 33$\%$.
Both the \FOH and \POH substrates were then optimized. In a next step, we considered
the \wc\ molecule in similar orientations for both classes of substrates and
 the corresponding geometries were then reoptimized with fixed in plane
lattice parameters but otherwise no symmetry constraints and the 
energies have been compared after relaxation. All calculations which involve the bonding of the precursor molecules
to the substrate were carried out in a 4 layered $3\times 3$ supercell
geometry in the plane perpendicular to the surface, corresponding to a
low coverage of precursor molecules in the initial growth stage.  The choice of the $3\times
3$ supercell geometry (with up to 256 atoms for the substrate/precursor
system) allows for a distance of more
than 8~{\AA} between two precursor molecules (when measured between
the two CO ligands of the adjacent molecules), thus providing an
optimal compromise between computational effort and minimization of
the interaction between the precursor molecules.  In all cases, the
distance between slabs (on the $c$ axis) was kept large enough 
(30 {\AA} that provides a distance of ~24 {\AA} on the substrate/precursor system)
to prevent significant interactions between the adjacent
images.

All calculations were performed in the scalar relativistic approximation.
A kinetic energy cut-off of 400~eV was used and all ions were fully
relaxed using a conjugate gradient scheme until the forces were less
than 0.01~eV/{\AA}. In the geometry optimizations for the molecule and the surface models
the Brillouin zone was sampled at the ${\Gamma}$ point only.
For the final comparison of energies and the density of states (DOS), a k-mesh of
$4\times 4\times 1$ has been used for the slabs showing a convergence
of the total energy to about $10^{-4}$~eV/f.u. For bulk {\sio} the {\bf k}-point
sampling was performed with a $16\times 16\times 16$ Monkhorst-Pack
grid. For all DOS calculations a Gaussian broadening of 20 meV was applied.
Spin polarization was considered for all calculations. Different spin states for each
cluster/adsorbate system were checked (i.e. in each case the two lowest
possible spin states) and only the results of the ground state are reported below.

In addition, the {\wc} molecule was optimized in the Oh symmetry in a cubic box of 30~{\AA} 
and the structural parameters were used to compare with the surface models. 
We also analyzed the electronic orbitals with Turbomole 6.0~\cite{Treutler1995,Eichkorn1997}
using the def-TZVP basis sets, the same GGA functional as for the bulk 
calculations as well as the RI (Resolution of Identity)
approximation. In this case, an effective core potential has been used for W, which provides 12 active valence
electrons.

The adsorption energy was defined as ${\Delta}E = E_{\rm total}-
E_{\rm substrate}- E_{\rm adsorbate}$, where $E_{\rm total}$, $E_{\rm
  substrate}$, and $E_{\rm adsorbate}$ are the energies of the
combined system (adsorbate and cluster), of the cluster, and of the
adsorbate molecule in the gas phase in a neutral state,
respectively.  

\section{Results}
\subsection{Electronic structure of the {\sio} substrate}

{\sio} exists in a variety of crystalline and amorphous
modifications.  In most experimental studies amorphous {\sio}
substrates are used for the growth process, which are however very
demanding for the computational simulations.  Therefore we use in our
study $\beta$-cristobalite, which resembles most closely amorphous
{\sio}, as a representative structure for amorphous {\sio}.
Experimental studies have shown that $\beta$-cristobalite and
amorphous {\sio} have a very similar local structure and exhibit
similar physical properties such as density and refractive
index~\cite{Jiang2005}. Following
Refs.~\onlinecite{Wehling2008,VigneMaeder1997,Ceresoli2000,Ricci2004}
we use the ideal {\bc} structure with  Fd\={3}m symmetry.

It has been reported that the silica surface may contain segments of
surfaces resembling both the (111) and (100) faces of
cristobalite~\cite{Peri1968,Sindorf1983}. The
Bravais-Donnay-Friedel-Harker (BDFH) method provides an estimate of
the relative growth rates of the possible faces of each crystal
structure and the resultant morphology, which shows that there exists
only one dominant plane for {{\sio}} which is
(111)~\cite{Puhakka2011}. Therefore, the surface of the substrate is
simulated by the (111) plane of bulk $\beta$-cristobalite. 

\begin{figure}
\includegraphics[width=0.5\textwidth]{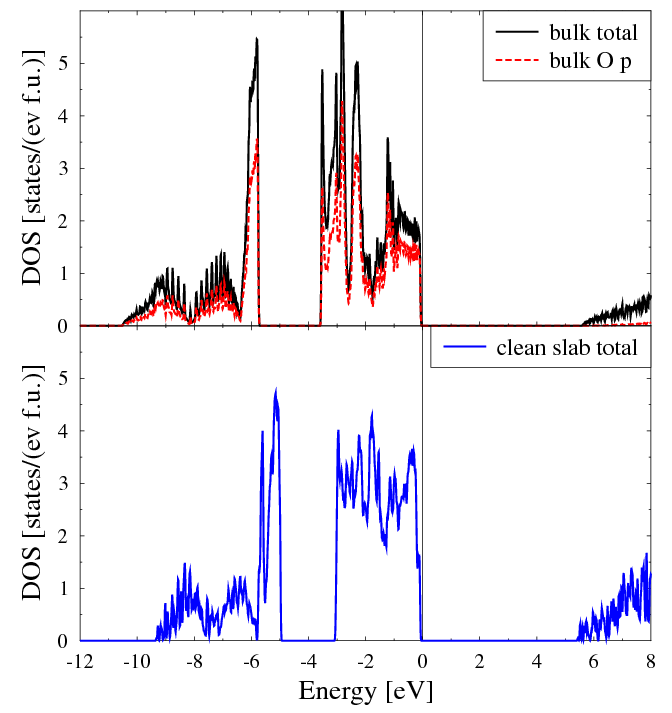}  
\caption{\label{fig:DOS_bulk_substrate} (Color online)
DOS of bulk {\sio} (top) in comparison to the OH-terminated 
4 layer slab (bottom). Shown are the respective total DOS (solid line)
and the partial O $2p$ contribution for the bulk (dashed line).
Results are shown for GGA calculations without vdW corrections.}
\end{figure}

For a slab
size of 4 layers we have observed a convergence in structural and
electronic properties.  In the optimized structure of 4 layer slabs
with no symmetry constraints the bond lengths of Si and O in the inner
two layers are found to be in the range of 1.62 - 1.63~{\AA}
matching well with the experimental values of
1.61~{\AA}~\cite{Wright1975} and previously reported theoretical
values~\cite{Yang2005}. Fig.~\ref{fig:DOS_bulk_substrate} shows the
DOS of bulk {\sio} (upper panel)
 and the DOS of the {\FOH} 4 layer slab (lower panel). The upper valence bands in the region
between -10 and 0~eV below the Fermi energy are dominated by O $2p$
states with some admixture of Si states (not shown).
The electronic structure of the 4 layer slab shows a reasonable convergence
with respect to the bulk (note that there are some smaller differences
due to finite size and relaxation effects as well as due to different 
{\bf k}-point sampling, which should however not affect the discussions
in the subsequent sections).

\subsection{Electronic structure of {\wc}  in the gas phase}
\begin{figure}  
\begin{center}  
\includegraphics[scale=0.25]{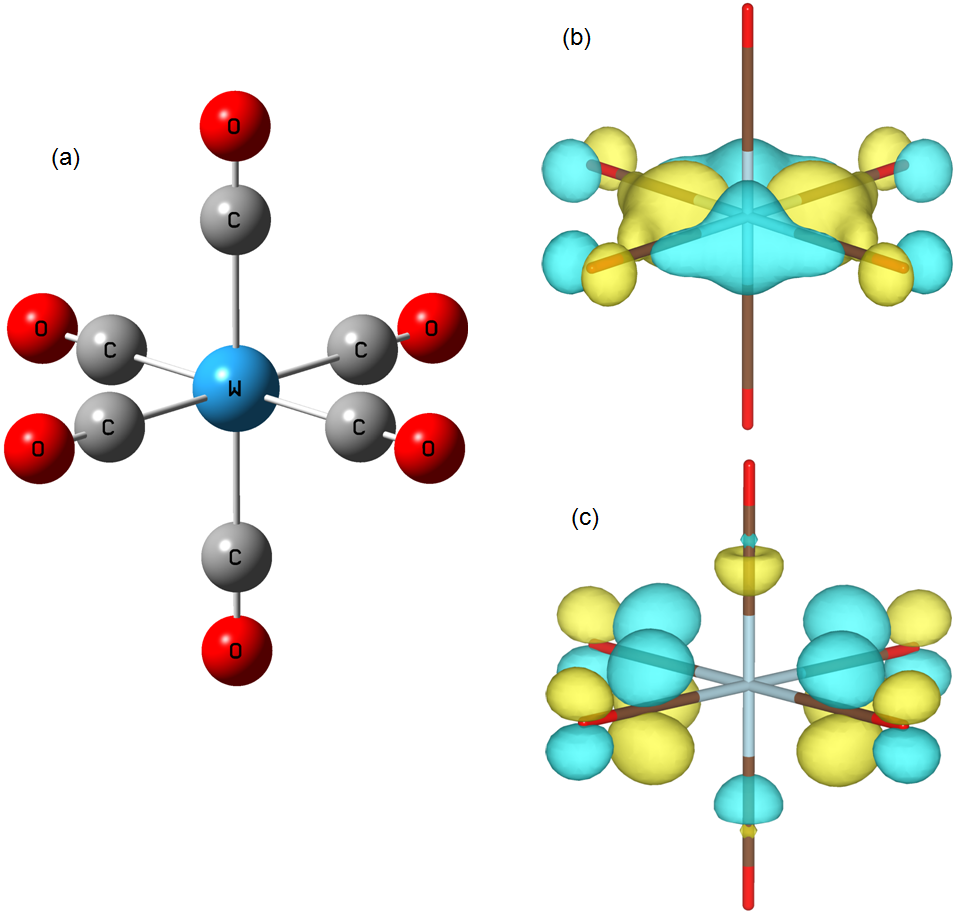}  
\caption{(Color online)
 (a) optimized structure of {\wc} and one of its triply degenerate (b) HOMO and (c)
  LUMO shown as a representative}\label{fig:WCO6}
\end{center}   
\end{figure}

The structural and electronic properties of {\wc} have been widely
investigated using local basis sets over the last two
decades~\cite{Rosa1999,Liu2010}. To compare the electronic and
energetic properties of {\wc} both in the gas phase and as an adsorbed
species, we have optimized the structure using a plane wave basis
set. The optimized structure is shown in Fig.~\ref{fig:WCO6}~(a).  The
calculated W-C bond length of the free molecule obtained using a plane
wave basis is found to be 2.06~{\AA} and the C-O bond length is
1.16~{\AA} compared to 2.06~{\AA}~\cite{Brockway1938} and
1.15~{\AA} respectively determined in experiment~\cite{Li1995}. Investigation
of the electronic structure of {\wc} shows that the highest occupied
state is dominated by W $5d$ orbitals. 
Indeed, the analysis of molecular orbitals also
confirms this fact.  The highest occupied molecular orbital (HOMO)
and the lowest unoccupied molecular orbital (LUMO) of \wc\ are shown in
Fig.~\ref{fig:WCO6}~(b) and (c), which show that the HOMO is made up of
W $5d$ while the LUMO is dominated by $p$ orbitals of the CO ligands.

\subsection{Interaction of {\wc} with the fully hydroxylated {\sio} substrate}
\begin{figure}  
\begin{center}  
\includegraphics[width=0.5\textwidth]{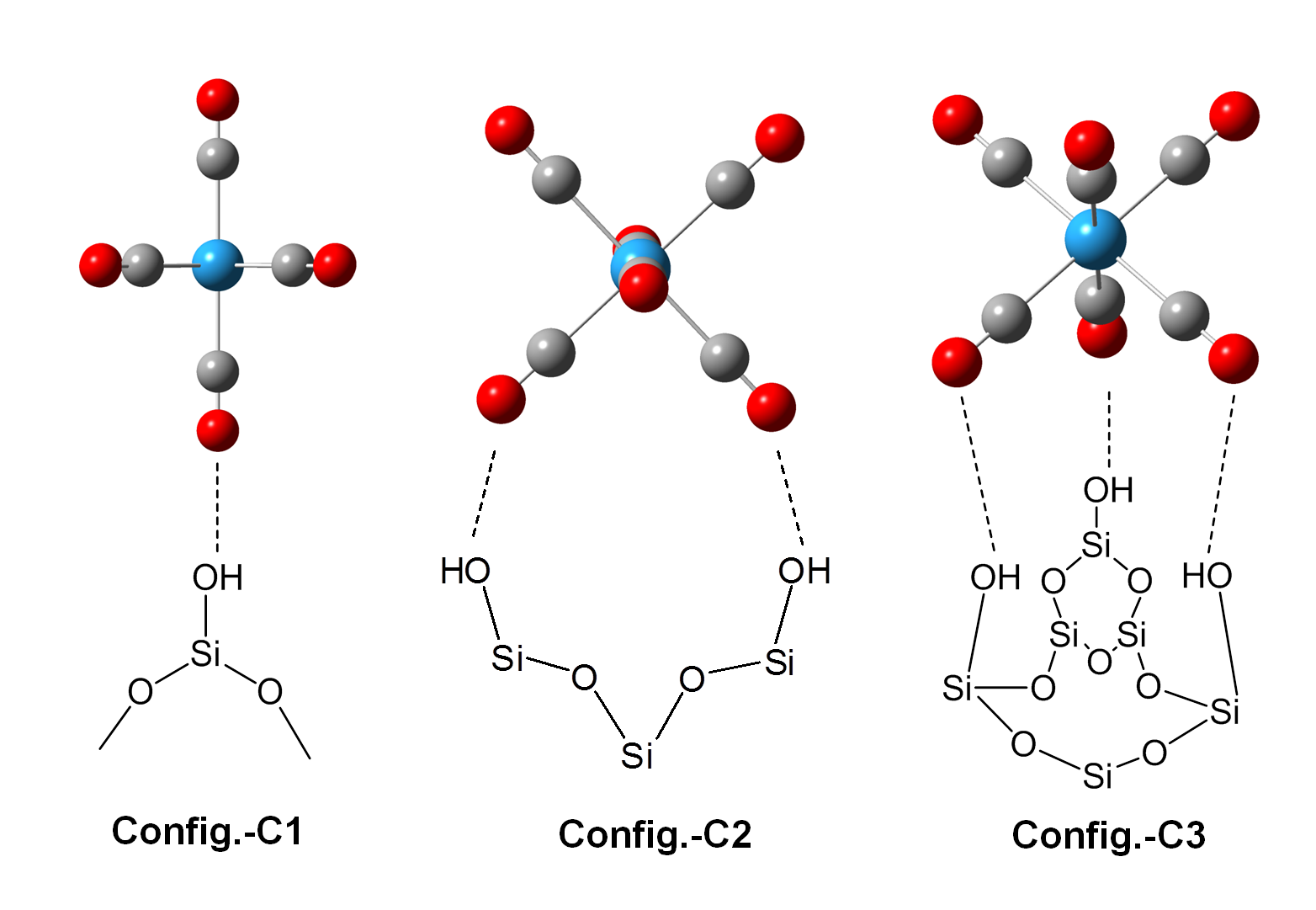} 
\includegraphics[width=0.48 \textwidth]{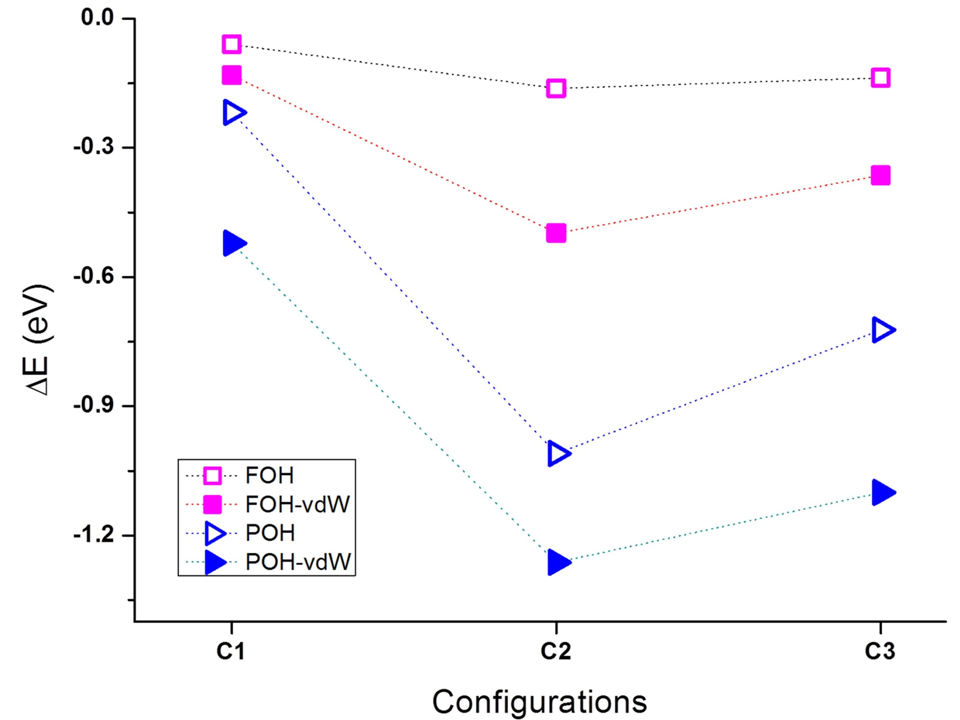}   
\caption{\label{fig:WCO6config} (Color online)
  Top panel: Three different configurations of {\wc}
  (C1, C2 and C3) considered for the adsorption
on the \FOH  surface
  are shown schematically. Similar configurations are considered on
  the \POH  surfaces where the CO ligands bond
  directly to the surface Si atoms. 
 Bottom panel: Variation of the adsorption energy 
${\Delta}E$ for the different configurations with respect to
  the orientation of {\wc}. Results are shown for the {\FOH} and {\POH} 
with and without inclusion of vdW corrections}
\end{center}   
\end{figure}

The \FOH surface corresponds to a substrate
prepared under wet chemical conditions in the absence of an electron
or ion beam.
In accordance with the available experimental results,
CO ligands of {\wc} have been used to bond to the
substrate~\cite{Myllyoja1999}.  Three different orientations are
considered which differ in the number of CO ligands bonding to
the surface. These configurations were subsequently relaxed 
without any symmetry constraints.
A schematic representation of the three configurations considered 
(C1, C2 and C3; C$n$ involves bonding to $n$ CO ligands) and
their corresponding adsorption energies ${\Delta}E$ are shown in
Fig.~\ref{fig:WCO6config}~(a) and (b).  

Our results illustrate that
the {\wc} molecule bonded with two of its CO ligands to the substrate (C2)
is the most stable configuration with an adsorption energy of -0.162~eV
upon neglect of vdW interactions.  
However, the energy difference compared to the next stable configuration, 
which involves three CO ligands bonding to the substrate (C3), is quite small
(about 25~meV), which may be indicative for a certain capacity of the 
molecule to roll over
the surface.  In the relaxed structure, the minimal distances observed
between the molecule CO groups and the surface OH groups are
2.07~{\AA} (C1), 2.14~{\AA} (C2) 
and 2.06~{\AA} for configuration C3.

The weak interaction between {\wc} and the \FOH
surface results only in minor rearrangements of the substrate surface
geometry. For instance, the Si-O bond distance on the adsorbate-free
surface is in the range of 1.62-1.65~{\AA}. For the
relaxed configuration with adsorbate we do not observe any significant
changes of these distances, illustrating the fact that the
surface does not contribute significantly to the bonding.  
We also observe only minor
changes in the structure of the {\wc} molecule bonded to the
surface.  In all the configurations we have considered, 
the W-C and the C-O bonds
that are involved in the bonding with the surface are shortened by
0.01-0.02~{\AA} which also confirms the weak adsorption.

\begin{figure}  
\begin{center}  
\includegraphics[width=0.48\textwidth]{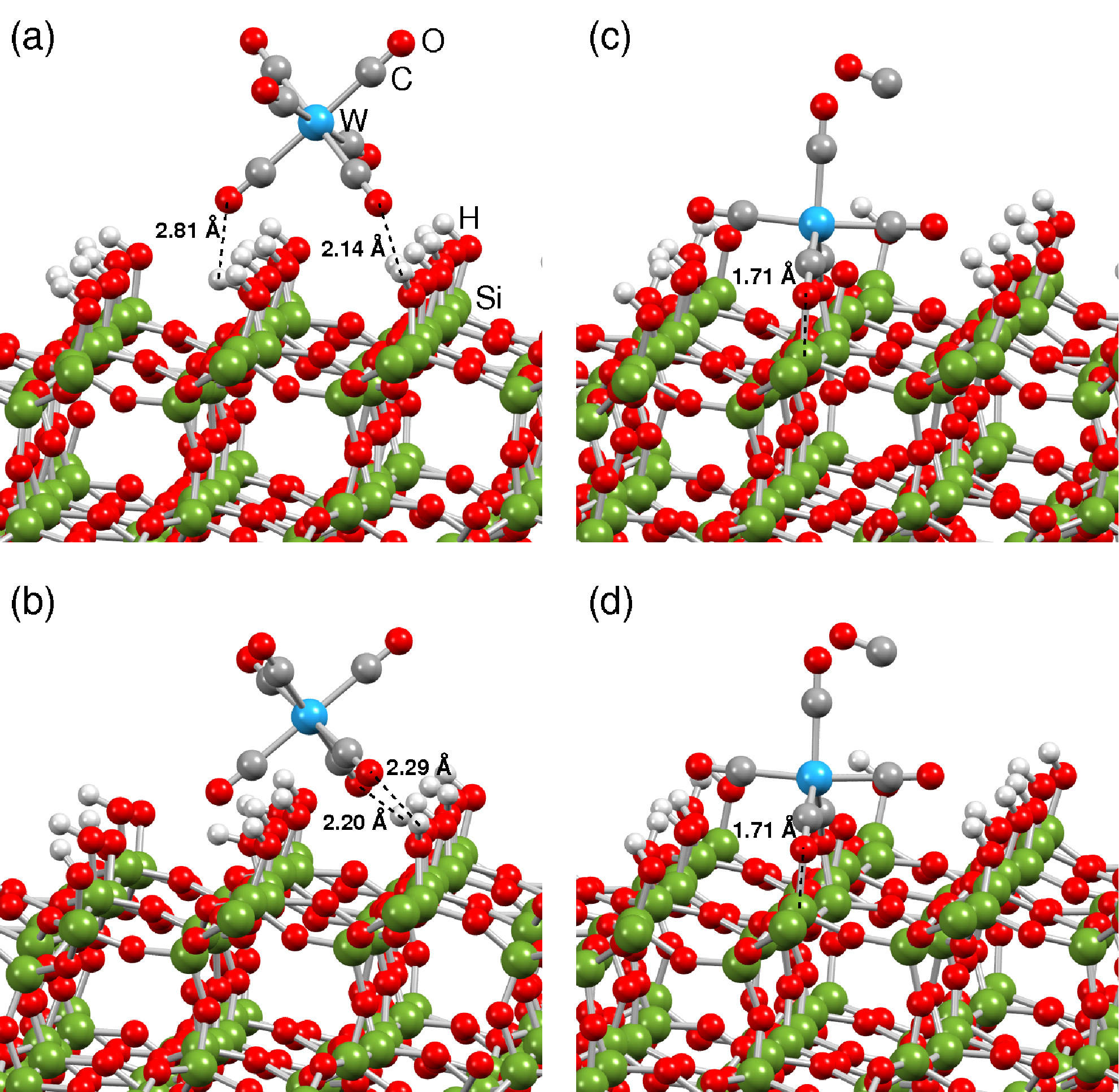}  
\caption{\label{fig:WCO6onSiO2} (Color online)
  Illustration of the role of the vdW
  correction in determining the bonding of {\wc}   (C2)
 to the {\FOH} ((a) and (b))  and
  {\POH} ((c) and (d)) surfaces. The upper panel ((a) and
(c)) correspond to calculations without vdW corrections while the
relaxed structures in the lower panel ((b) and (d)) include vdW corrections.}
\end{center}   
\end{figure}

The role of vdW forces on these adsorption geometries is
evaluated by adding vdW corrections to the calculations which
are not accounted for by DFT within the GGA functional.  
As expected, the absolute values of the adsorption energies are increased
for all probed configurations~(Fig.~\ref{fig:WCO6config} (b)). For the
most stable configuration (C2) we find an  adsorption energy 
$\Delta E= -0.498$ eV.

We observe
minor changes in the structural parameters when vdW corrections
are included in the calculations compared to the set of calculations
without these corrections.  For example, the minimal distance between
the surface and the molecule in the three configurations are 2.11~{\AA}
(C1), 2.20~{\AA} (C2) and 2.08~{\AA} (C3) and the
changes in W-C and C-O bond lengths are very small (of the order of
0.01~{\AA}) compared to the calculations without dispersion
corrections. 

Although the minimal distance between the molecule and
the surface are similar in both cases (C2 and C3), for the preferred configuration (C2),
 the remaining CO ligands are found to lie closer to the surface 
with the next nearest distances 
of 2.29~{\AA} and
2.65~{\AA} (see Fig.~\ref{fig:WCO6onSiO2} (b)) 
compared to 2.82~{\AA} and 3.43~{\AA} (see Fig.~\ref{fig:WCO6onSiO2} (a)) when the
corrections are not implemented.  This in fact has an impact on the
orientation of the molecule on the surface. Furthermore, the differences
between the adsorption energies of the different configurations are now
larger, thus indicating that vdW interactions tend to decrease the ability
of the precursor molecule to spontaneously roll over the
surface.

\begin{figure}  
\begin{center}  
\includegraphics[width=0.5\textwidth]{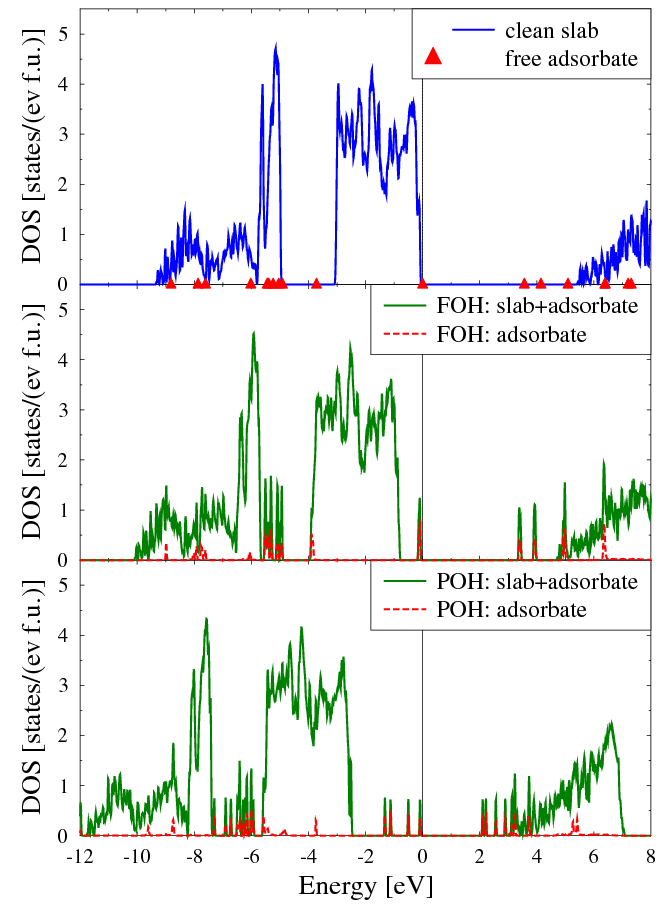} 
\caption{\label{fig:WCO6DOS} (Color online)
DOS of {\wc} on {\sio} in comparison to the 
clean {\sio} surface. Shown are the total DOS (solid lines) along with the
partial contributions of the adsorbate (dashed lines) for the preferential 
configurations on the {\POH} surface (bottom panel) and
on the {\FOH} surface (middle panel).
The top panel shows the total DOS of the {\sio} surface (solid line)
along with the energetic levels of the free {\wc} molecule.
The Fermi energy is set to zero.
} 
\end{center}   
\end{figure}

The DOS of  {\wc} on the \FOH 
surface for C2 is shown in Fig.~\ref{fig:WCO6DOS} (middle panel)
in comparison to the DOS of the slab without adsorbate (top panel).
Results are shown for calculations including vdW corrections, which however
have only minor influence on the electronic structure (compare 
Fig.~\ref{fig:DOS_bulk_substrate} (bottom panel)). 
The highest molecular level of the
W(CO)$_6$ adsorbate lies in the gap of the substrate, 
which pins the Fermi level close to the valence band.
Thus the  DOS of the {\sio} substrate consisting
mostly of O $p$ states is shifted to lower
energies with respect to the Fermi energy. 
Otherwise the  DOS of the {\sio} substrate shows only minor 
modifications with respect to the free substrate.
The molecular levels of the {\wc} adsorbate are visible in sharp peaks 
and agree well with the energetic levels of the free {\wc} molecule
obtained by our Turbomole calculations (also shown in the top panel
of Fig.~\ref{fig:WCO6DOS}).
This further illustrates the weak interaction between the
adsorbate molecule and the {\sio} substrate.

\subsection{Interaction of {\wc} with the partially hydroxylated {\sio} substrate}\label{sec3.4}

Partial de-hydroxylation of the surface of a wet chemically prepared
{\sio} substrate is expected to occur under irradiation with an
electron or ion beam or at elevated temperatures. To represent this
situation in this study, we kept the Si atoms bonding to CO ligands
unterminated, while all other dangling bonds of Si on the two surfaces
of the slab were terminated by OH groups.  The {\wc} molecule was
placed in such a way that it bonds with one, two or three CO ligands
to the unterminated Si surface atom and the substrate/precursor system
was subsequently relaxed.

Our results indicate that the {\wc} molecule is adsorbed more strongly on the
\POH surface in all three orientations compared to the
\FOH surface (see Fig.~\ref{fig:WCO6config}~(b)). Also
on the \POH surface the precursor
molecule prefers to be bonded through multiple CO ligands. In the multiple
bonding configurations (C2 and C3) we observe a spontaneous
fragmentation of the {\wc} precursor molecule with chemisorption
of a W(CO)$_5$ fragment on the surface upon removal of one of the CO
ligands. 

The most stable configuration (C2) after relaxation with $\Delta E= -1.010$ 
eV upon neglect of vdW interactions is
shown in Fig.~\ref{fig:WCO6onSiO2}~(c). 
The fragmented CO lies at a distance of 4.15 \AA\ from the sub-carbonyl moiety
(measured between 'W' and CO) and 5.03 \AA\ from the surface (measured between C and nearest H atom of the surface OH group)
and remains as a free CO molecule.
The bond distances between Si and the O atoms of the CO ligands involved in the bonding are about
1.71~{\AA} and quite close to the Si-O bond lengths of 1.62 - 1.65~{\AA} observed in various
compounds~\cite{Baur1978}, indicating a nearly optimal interaction.
The distances between the O of the two non-bonding CO ligands 
to the H atoms of the OH group are 1.94 and 1.98~{\AA} indicating
the presence of a weak interaction which might also stabilize the molecule
on the surface. 

A similar effect is also seen for the singly coordinated system (C1).
When the {\wc} molecule is bonded by only one of its CO ligands
to the \POH surface, the distance between Si and O of the nearest 
CO ligand is 1.73~{\AA} with the fairly small adsorption energy $\Delta E=-0.218$~eV. 
The distance of non-bonding CO ligands to the substrate are 2.31 and 2.58~{\AA}
compared to 2.08 - 2.48~{\AA} found for \FOH surfaces. 
In the three coordinated system (C3), where the {\wc} molecule is bonded by three of
its CO ligands to the \POH surface, the Si-O (of CO) distances range from 
1.69 - 1.70~{\AA} with an adsorption energy $\Delta E$ of -0.723~eV.

In the preferred configuration (C2) the W-C bond lengths of
1.87$-$1.94~{\AA} for the CO groups which are involved in the
bonding to the substrate are shorter compared to the value found in
the gas phase 2.05~{\AA}. On the other hand side, the bond lengths of 
the non-bonding CO ligands are slightly elongated (2.11 - 2.12~{\AA}), illustrating a
weakening of these bonds.  

Calculations have also been carried out by including  vdW
corrections.  No major changes are observed in the structural parameters
of {\wc}, (see also Fig.~\ref{fig:WCO6onSiO2}~(c) and (d)). However,
differences are observed for the calculated $\Delta E$ 
on these \POH  surfaces (Fig.~\ref{fig:WCO6config}~(b)). 
The most stable configuration (C2) has an adsorption energy 
$\Delta E = -1.262$ eV when vdW interactions are included.
Moreover, the differences in $\Delta E$ between bi
and tri -coordinate orientation of {\wc} on these surfaces becomes
small, illustrating the presence of
physisorption although dominated by chemisorption in these
cases. Irrespective of either the orientation or the inclusion of vdW
corrections, fragmentation of {\wc} molecules is found to occur.

\begin{figure}  
\begin{center}  
\includegraphics[width=0.5\textwidth]{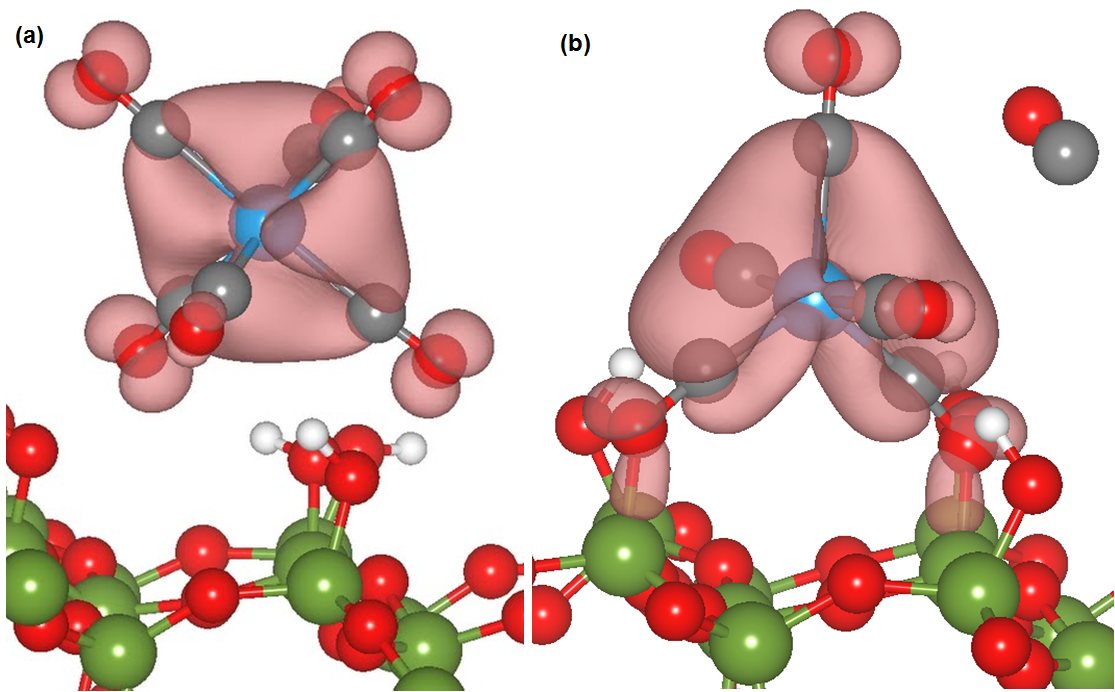} 
\caption{\label{fig:bWCO6DOS} (Color online)
Band decomposed charge density for the valence band maximum
for the {\FOH} (a) and {\POH} (b) cases.} 
\end{center}   
\end{figure}

The DOS of the preferential configuration of {\wc} on the \POH surface (C2) is shown in the bottom panel of
Fig.~\ref{fig:WCO6DOS}. Similar to the case of {\wc} on the \FOH surface,
the overall shape of the partial DOS of the substrate 
is only little altered by the interaction with the precursor molecule, but 
shifted to lower energies with respect to the Fermi level since the highest
occupied level of the adsorbate lies in the gap of the substrate.
However, in contrast to the result for the \FOH
surface, the energetic levels of the adsorbate now have little in common
with those of the free gas phase {\wc} molecule, as a consequence of the
stronger interaction and the fragmentation of the precursor molecule.

In order to analyze the nature of the bonding between precursor
and SiO$_2$ substrate in the adsorption process, we show in Fig.~\ref{fig:bWCO6DOS}
the charge density of the highest occupied valence band 
of {\wc} adsorbed on \FOH (see Fig.~\ref{fig:bWCO6DOS} (a)) and \POH (see Fig.~\ref{fig:bWCO6DOS} (b))
 surfaces. 
Our results indicate that the degeneracy of the W $5d$ 
triply degenerate HOMO level is nearly
preserved on the \FOH surface (cf. Figs.~\ref{fig:WCO6} (b) and~\ref{fig:bWCO6DOS} (a)), while it is strongly altered as a consequence of
the symmetry lowering while binding to the \POH substrate (cf. Figs..~\ref{fig:WCO6} (b) and~\ref{fig:bWCO6DOS} (b)).
This observation illustrates the fact that the electron density on the tungsten
is perturbed upon bonding to dehydroxylated Si atoms located on the {\sio} surface which
acts as an  electron accepting centre \cite{morrow}. Our computed Bader charges on
the  \POH surface before and after bonding illustrate that the dehydroxylated Si atom on the substrate gain
0.9 electrons as a result of bonding.
This interaction results in the formation of a ligand-surface
adduct which destabilises the molecule by weakening the W-CO bonds that were not involved in bonding to the substrate and
ultimately eliminates one of its  CO ligands. 
In fact, this behaviour is similar to the interaction of group VI metal carbonyls with the
 Al$^{3+}$ sites
of ${\gamma}$-{Al$_2$O$_3$} resulting in the formation of subcarbonyl
 entities bonded to the surface. \cite{adriano}
Owing to the high adsorption energy, the  formed sub-carbonyl moiety has
 less mobility on the surface
than the unfragmented molecule. In summary,
 the observation of high adsorption energy of {W(CO)$_5$}, the
elongation of the W-CO bonds oriented towards the vacuum and the fragmentation of
one of the CO ligands on the \POH surface confirms that {\wc} bonds in a similar 
fashion in \POH the surface as on ${\gamma}$-{Al$_2$O$_3$}.
  
\section{Discussion: Towards a realistic simulation of the growth process}

The results obtained in the previous section show that {\wc} is
only weakly adsorbed on a \FOH surface. The calculated
adsorption energies in the range between $\Delta E\approx$ -0.1 to -0.5 eV 
may eventually allow the growth of {\wc} layers at low
temperatures, but do not lead to the growth of metallic W
deposits. Furthermore, the small energy differences between the probed
configurations of {\wc} on the {\FOH} substrate are indicative
of a high mobility on the surface, although the energy barriers
between the configurations are hard to estimate. Experimentally, the
formation of {\wc} layers has been observed by exposure of a
TiO$_2$(110) surface to {\wc} molecules at T=140 K, which however
desorb again after annealing to room
temperature~\cite{Domenichini2008}.
On the contrary, we observe chemisorption of {\wc} with a
spontaneous fragmentation of the precursor molecule into W(CO)$_5$ upon
removal of a volatile CO ligand on a \POH surface. 
The possibility of mobility on the \FOH surfaces and fragmentation of the CO ligands on the
\POH surfaces observed in this study illustrates that
the molecule may roll over to the {\POH} surface regions and
get activated for fragmentation.

The rate limiting step in the decomposition of {\wc} is believed to be the
elimination of the first W$-$CO bond. In the gas phase, this elimination
requires about 45 kcal/mol~\cite{Lewis1984,Sniatynsky2004}, but is
expected to be essentially smaller on the substrate and has been
estimated to be about 12 kcal/mol for {\wc} on Ni(100)~\cite{Zaera1992}. 
Further, experiments have found that complete de-carbonylation occurs at around $150-205\,^{\circ}\mathrm{C}$ in the
CVD experiments~\cite{Kiminami2003,Zaera1992}. 
This in fact, is supported by the weakening of bonds of the CO ligands not involved in the bonding to
the substrate observed in our calculations, which simplifies the elimination of further ligands by thermal or radiative processes. Our
results thus show that the spontaneous fragmentation of {\wc} on a {\POH} surface is a relevant process to be taken
into account for the simulation of the initial stage of the growth of tungsten nano-deposits.

\section{CONCLUSIONS}\label{Summarize}
The interaction between a {\wc} precursor molecule and {\sio}
substrates has been studied in the framework of density functional
theory.  
Our results show that the interaction between the
precursor molecule and the \FOH substrate results in a
weak adsorption of the precursor molecule.  
However, upon partial de-hydroxylation of the surface, we
observe a chemisorption and fragmentation of the precursor molecule,
which is explained in terms of the changes of the electronic
properties and the charge transfer between the precursor molecule and
the substrate.
We observe that vdW corrections result in a re-orientation
and stabilization of the molecules on the \FOH surface,
but have only a minor impact on the adsorption geometry for the
\POH surface.
  
Our reported calculations provide insight into  various aspects of the interaction of the
precursor molecule with the {\sio} substrate.  In a real growth of nano-deposits as is the case of 
 electron beam induced deposition of precursor molecules on a substrate, 
 there are a multitude of processes to be taken into account. The interaction of
the beam with the precursor molecules in the gas phase for instance
will lead to an activation or fragmentation of the precursor
molecules.  However, since in focused EBID experiments the fraction of
gas phase molecules interacting directly with the beam is small, this
process may be expected to be of minor relevance for
EBID~\cite{HuthPrivate}.  Furthermore, the
interaction of the beam with the substrate will lead to heating
 and charging of the substrate as well as to emission of secondary
electrons and photons.  These processes can be partially taken into
account by finite temperature molecular dynamics simulations, which we plan to do
 in the future but are  beyond the scope
of the present investigations.

\section{Acknowledgments}\label{Acknowledgments}
The authors gratefully acknowledge financial support by the
Beilstein-Institut, Frankfurt/Main, Germany, within the research
collaboration NanoBiC. This work was supported by the Alliance Program
of the Helmholtz Association (HA216/EMMI).  The generous allotment of
computer time by CSC-Frankfurt and LOEWE-CSC is gratefully acknowledged.


\begin{thebibliography}{52}%
\makeatletter
\providecommand \@ifxundefined [1]{%
 \@ifx{#1\undefined}
}%
\providecommand \@ifnum [1]{%
 \ifnum #1\expandafter \@firstoftwo
 \else \expandafter \@secondoftwo
 \fi
}%
\providecommand \@ifx [1]{%
 \ifx #1\expandafter \@firstoftwo
 \else \expandafter \@secondoftwo
 \fi
}%
\providecommand \natexlab [1]{#1}%
\providecommand \enquote  [1]{``#1''}%
\providecommand \bibnamefont  [1]{#1}%
\providecommand \bibfnamefont [1]{#1}%
\providecommand \citenamefont [1]{#1}%
\providecommand \href@noop [0]{\@secondoftwo}%
\providecommand \href [0]{\begingroup \@sanitize@url \@href}%
\providecommand \@href[1]{\@@startlink{#1}\@@href}%
\providecommand \@@href[1]{\endgroup#1\@@endlink}%
\providecommand \@sanitize@url [0]{\catcode `\\12\catcode `\$12\catcode
  `\&12\catcode `\#12\catcode `\^12\catcode `\_12\catcode `\%12\relax}%
\providecommand \@@startlink[1]{}%
\providecommand \@@endlink[0]{}%
\providecommand \url  [0]{\begingroup\@sanitize@url \@url }%
\providecommand \@url [1]{\endgroup\@href {#1}{\urlprefix }}%
\providecommand \urlprefix  [0]{URL }%
\providecommand \Eprint [0]{\href }%
\providecommand \doibase [0]{http://dx.doi.org/}%
\providecommand \selectlanguage [0]{\@gobble}%
\providecommand \bibinfo  [0]{\@secondoftwo}%
\providecommand \bibfield  [0]{\@secondoftwo}%
\providecommand \translation [1]{[#1]}%
\providecommand \BibitemOpen [0]{}%
\providecommand \bibitemStop [0]{}%
\providecommand \bibitemNoStop [0]{.\EOS\space}%
\providecommand \EOS [0]{\spacefactor3000\relax}%
\providecommand \BibitemShut  [1]{\csname bibitem#1\endcsname}%
\let\auto@bib@innerbib\@empty
\bibitem [{\citenamefont {Wnuk}\ \emph {et~al.}(2011)\citenamefont {Wnuk},
  \citenamefont {Rosenberg}, \citenamefont {Gorham}, \citenamefont {van Dorp},
  \citenamefont {Hagen},\ and\ \citenamefont {Fairbrother}}]{Wnuk2011}%
  \BibitemOpen
  \bibfield  {author} {\bibinfo {author} {\bibfnamefont {J.~D.}\ \bibnamefont
  {Wnuk}}, \bibinfo {author} {\bibfnamefont {S.~G.}\ \bibnamefont {Rosenberg}},
  \bibinfo {author} {\bibfnamefont {J.~M.}\ \bibnamefont {Gorham}}, \bibinfo
  {author} {\bibfnamefont {W.~F.}\ \bibnamefont {van Dorp}}, \bibinfo {author}
  {\bibfnamefont {C.~W.}\ \bibnamefont {Hagen}}, \ and\ \bibinfo {author}
  {\bibfnamefont {D.~H.}\ \bibnamefont {Fairbrother}},\ }\href {\doibase DOI:
  10.1016/j.susc.2010.10.035} {\bibfield  {journal} {\bibinfo  {journal} {Surf.
  Sci.}\ }\textbf {\bibinfo {volume} {605}},\ \bibinfo {pages} {257 } (\bibinfo
  {year} {2011})}\BibitemShut {NoStop}%
\bibitem [{\citenamefont {Utke}\ \emph {et~al.}(2008)\citenamefont {Utke},
  \citenamefont {Hoffmann},\ and\ \citenamefont {Melngailis}}]{Utke2008}%
  \BibitemOpen
  \bibfield  {author} {\bibinfo {author} {\bibfnamefont {I.}~\bibnamefont
  {Utke}}, \bibinfo {author} {\bibfnamefont {P.}~\bibnamefont {Hoffmann}}, \
  and\ \bibinfo {author} {\bibfnamefont {J.}~\bibnamefont {Melngailis}},\
  }\href {\doibase 10.1116/1.2955728} {\bibfield  {journal} {\bibinfo
  {journal} {J. Vac. Sci. Technol. B}\ }\textbf {\bibinfo {volume} {26}},\
  \bibinfo {pages} {1197} (\bibinfo {year} {2008})}\BibitemShut {NoStop}%
\bibitem [{\citenamefont {Arumainayagam}\ \emph {et~al.}(2010)\citenamefont
  {Arumainayagam}, \citenamefont {Lee}, \citenamefont {Nelson}, \citenamefont
  {Haines},\ and\ \citenamefont {Gunawardane}}]{Arumainayagam2010}%
  \BibitemOpen
  \bibfield  {author} {\bibinfo {author} {\bibfnamefont {C.~R.}\ \bibnamefont
  {Arumainayagam}}, \bibinfo {author} {\bibfnamefont {H.-L.}\ \bibnamefont
  {Lee}}, \bibinfo {author} {\bibfnamefont {R.~B.}\ \bibnamefont {Nelson}},
  \bibinfo {author} {\bibfnamefont {D.~R.}\ \bibnamefont {Haines}}, \ and\
  \bibinfo {author} {\bibfnamefont {R.~P.}\ \bibnamefont {Gunawardane}},\
  }\href {\doibase 10.1016/j.surfrep.2009.09.001} {\bibfield  {journal}
  {\bibinfo  {journal} {Surface Science Reports}\ }\textbf {\bibinfo {volume}
  {65}},\ \bibinfo {pages} {1 } (\bibinfo {year} {2010})}\BibitemShut {NoStop}%
\bibitem [{\citenamefont {Seuret}\ \emph {et~al.}(2003)\citenamefont {Seuret},
  \citenamefont {Cicoira}, \citenamefont {Ohta}, \citenamefont {Doppelt},
  \citenamefont {Hoffmann}, \citenamefont {Weber},\ and\ \citenamefont
  {Wesolowski}}]{Seuret2003}%
  \BibitemOpen
  \bibfield  {author} {\bibinfo {author} {\bibfnamefont {P.}~\bibnamefont
  {Seuret}}, \bibinfo {author} {\bibfnamefont {F.}~\bibnamefont {Cicoira}},
  \bibinfo {author} {\bibfnamefont {T.}~\bibnamefont {Ohta}}, \bibinfo {author}
  {\bibfnamefont {P.}~\bibnamefont {Doppelt}}, \bibinfo {author} {\bibfnamefont
  {P.}~\bibnamefont {Hoffmann}}, \bibinfo {author} {\bibfnamefont
  {J.}~\bibnamefont {Weber}}, \ and\ \bibinfo {author} {\bibfnamefont {T.~A.}\
  \bibnamefont {Wesolowski}},\ }\href {\doibase 10.1039/B206731E} {\bibfield
  {journal} {\bibinfo  {journal} {Phys. Chem. Chem. Phys.}\ }\textbf {\bibinfo
  {volume} {5}},\ \bibinfo {pages} {268} (\bibinfo {year} {2003})}\BibitemShut
  {NoStop}%
\bibitem [{\citenamefont {Guillam\'on}\ \emph {et~al.}(2008)\citenamefont
  {Guillam\'on}, \citenamefont {Suderow}, \citenamefont {Vieira}, \citenamefont
  {Fern\'andez-Pacheco}, \citenamefont {Ses\'e}, \citenamefont {C\'ordoba},
  \citenamefont {De~Teresa},\ and\ \citenamefont {Ibarra}}]{Guillamon2008}%
  \BibitemOpen
  \bibfield  {author} {\bibinfo {author} {\bibfnamefont {I.}~\bibnamefont
  {Guillam\'on}}, \bibinfo {author} {\bibfnamefont {H.}~\bibnamefont
  {Suderow}}, \bibinfo {author} {\bibfnamefont {S.}~\bibnamefont {Vieira}},
  \bibinfo {author} {\bibfnamefont {A.}~\bibnamefont {Fern\'andez-Pacheco}},
  \bibinfo {author} {\bibfnamefont {J.}~\bibnamefont {Ses\'e}}, \bibinfo
  {author} {\bibfnamefont {R.}~\bibnamefont {C\'ordoba}}, \bibinfo {author}
  {\bibfnamefont {J.~M.}\ \bibnamefont {De~Teresa}}, \ and\ \bibinfo {author}
  {\bibfnamefont {M.~R.}\ \bibnamefont {Ibarra}},\ }\href {\doibase
  10.1088/1367-2630/10/9/093005} {\bibfield  {journal} {\bibinfo  {journal}
  {New J. Phys.}\ }\textbf {\bibinfo {volume} {10}},\ \bibinfo {pages} {093005}
  (\bibinfo {year} {2008})}\BibitemShut {NoStop}%
\bibitem [{\citenamefont {Koops}\ \emph {et~al.}(1988)\citenamefont {Koops},
  \citenamefont {Weiel}, \citenamefont {Kern},\ and\ \citenamefont
  {Baum}}]{Koops1988}%
  \BibitemOpen
  \bibfield  {author} {\bibinfo {author} {\bibfnamefont {H.~W.~P.}\
  \bibnamefont {Koops}}, \bibinfo {author} {\bibfnamefont {R.}~\bibnamefont
  {Weiel}}, \bibinfo {author} {\bibfnamefont {D.~P.}\ \bibnamefont {Kern}}, \
  and\ \bibinfo {author} {\bibfnamefont {T.~H.}\ \bibnamefont {Baum}},\ }\href
  {\doibase 10.1116/1.584045} {\bibfield  {journal} {\bibinfo  {journal} {J.
  Vac. Sci. Technol. B}\ }\textbf {\bibinfo {volume} {6}},\ \bibinfo {pages}
  {477} (\bibinfo {year} {1988})}\BibitemShut {NoStop}%
\bibitem [{\citenamefont {Okuyama}(1980)}]{Okuyama1980}%
  \BibitemOpen
  \bibfield  {author} {\bibinfo {author} {\bibfnamefont {F.}~\bibnamefont
  {Okuyama}},\ }\href {\doibase 10.1063/1.91310} {\bibfield  {journal}
  {\bibinfo  {journal} {Appl. Phys. Lett.}\ }\textbf {\bibinfo {volume} {36}},\
  \bibinfo {pages} {46} (\bibinfo {year} {1980})}\BibitemShut {NoStop}%
\bibitem [{\citenamefont {Flitsch}\ \emph {et~al.}(1991)\citenamefont
  {Flitsch}, \citenamefont {Swanson},\ and\ \citenamefont
  {Friend}}]{Flitsch1991}%
  \BibitemOpen
  \bibfield  {author} {\bibinfo {author} {\bibfnamefont {F.~A.}\ \bibnamefont
  {Flitsch}}, \bibinfo {author} {\bibfnamefont {J.~R.}\ \bibnamefont
  {Swanson}}, \ and\ \bibinfo {author} {\bibfnamefont {C.~M.}\ \bibnamefont
  {Friend}},\ }\href {\doibase 10.1016/0039-6028(91)90470-D} {\bibfield
  {journal} {\bibinfo  {journal} {Surf. Sci.}\ }\textbf {\bibinfo {volume}
  {245}},\ \bibinfo {pages} {85 } (\bibinfo {year} {1991})}\BibitemShut
  {NoStop}%
\bibitem [{\citenamefont {Zaera}(1992)}]{Zaera1992}%
  \BibitemOpen
  \bibfield  {author} {\bibinfo {author} {\bibfnamefont {F.}~\bibnamefont
  {Zaera}},\ }\href {\doibase 10.1021/j100190a086} {\bibfield  {journal}
  {\bibinfo  {journal} {J. Phys. Chem.}\ }\textbf {\bibinfo {volume} {96}},\
  \bibinfo {pages} {4609} (\bibinfo {year} {1992})}\BibitemShut {NoStop}%
\bibitem [{\citenamefont {Brenner}\ \emph {et~al.}(1979)\citenamefont
  {Brenner}, \citenamefont {Hucul},\ and\ \citenamefont
  {Hardwick}}]{Brenner1979}%
  \BibitemOpen
  \bibfield  {author} {\bibinfo {author} {\bibfnamefont {A.}~\bibnamefont
  {Brenner}}, \bibinfo {author} {\bibfnamefont {D.~A.}\ \bibnamefont {Hucul}},
  \ and\ \bibinfo {author} {\bibfnamefont {S.~J.}\ \bibnamefont {Hardwick}},\
  }\href {\doibase 10.1021/ic50196a015} {\bibfield  {journal} {\bibinfo
  {journal} {Inorg. Chem.}\ }\textbf {\bibinfo {volume} {18}},\ \bibinfo
  {pages} {1478} (\bibinfo {year} {1979})}\BibitemShut {NoStop}%
\bibitem [{\citenamefont {Linden}\ \emph {et~al.}(1980)\citenamefont {Linden},
  \citenamefont {Beckey},\ and\ \citenamefont {Okuyama}}]{Linden1980}%
  \BibitemOpen
  \bibfield  {author} {\bibinfo {author} {\bibfnamefont {H.~B.}\ \bibnamefont
  {Linden}}, \bibinfo {author} {\bibfnamefont {H.~D.}\ \bibnamefont {Beckey}},
  \ and\ \bibinfo {author} {\bibfnamefont {F.}~\bibnamefont {Okuyama}},\ }\href
  {http://dx.doi.org/10.1007/BF00897937} {\bibfield  {journal} {\bibinfo
  {journal} {Appl. Phys.}\ }\textbf {\bibinfo {volume} {22}},\ \bibinfo {pages}
  {83} (\bibinfo {year} {1980})}\BibitemShut {NoStop}%
\bibitem [{\citenamefont {Kuzusaka}\ and\ \citenamefont
  {Howe}(1980)}]{Kuzusaka1980}%
  \BibitemOpen
  \bibfield  {author} {\bibinfo {author} {\bibfnamefont {A.}~\bibnamefont
  {Kuzusaka}}\ and\ \bibinfo {author} {\bibfnamefont {R.~F.}\ \bibnamefont
  {Howe}},\ }\href {\doibase 10.1016/0304-5102(80)80006-X} {\bibfield
  {journal} {\bibinfo  {journal} {Journal of Molecular Catalysis}\ }\textbf
  {\bibinfo {volume} {9}},\ \bibinfo {pages} {199 } (\bibinfo {year}
  {1980})}\BibitemShut {NoStop}%
\bibitem [{\citenamefont {Ehlers}\ and\ \citenamefont
  {Frenking}(1994)}]{Ehlers1994}%
  \BibitemOpen
  \bibfield  {author} {\bibinfo {author} {\bibfnamefont {A.~W.}\ \bibnamefont
  {Ehlers}}\ and\ \bibinfo {author} {\bibfnamefont {G.}~\bibnamefont
  {Frenking}},\ }\href {\doibase 10.1021/ja00083a040} {\bibfield  {journal}
  {\bibinfo  {journal} {J. Am. Chem. Soc.}\ }\textbf {\bibinfo {volume}
  {116}},\ \bibinfo {pages} {1514} (\bibinfo {year} {1994})}\BibitemShut
  {NoStop}%
\bibitem [{\citenamefont {Kurhinen}\ \emph {et~al.}(1994)\citenamefont
  {Kurhinen}, \citenamefont {Venalainen},\ and\ \citenamefont
  {Pakkanen}}]{Kurhinen1994}%
  \BibitemOpen
  \bibfield  {author} {\bibinfo {author} {\bibfnamefont {M.}~\bibnamefont
  {Kurhinen}}, \bibinfo {author} {\bibfnamefont {T.}~\bibnamefont
  {Venalainen}}, \ and\ \bibinfo {author} {\bibfnamefont {T.~A.}\ \bibnamefont
  {Pakkanen}},\ }\href {\doibase 10.1021/j100091a045} {\bibfield  {journal}
  {\bibinfo  {journal} {J. Phys. Chem.}\ }\textbf {\bibinfo {volume} {98}},\
  \bibinfo {pages} {10237} (\bibinfo {year} {1994})}\BibitemShut {NoStop}%
\bibitem [{\citenamefont {Suvanto}\ and\ \citenamefont
  {Pakkanen}(1999)}]{Suvanto1999}%
  \BibitemOpen
  \bibfield  {author} {\bibinfo {author} {\bibfnamefont {M.}~\bibnamefont
  {Suvanto}}\ and\ \bibinfo {author} {\bibfnamefont {T.~A.}\ \bibnamefont
  {Pakkanen}},\ }\href {\doibase 10.1016/S1381-1169(98)00138-1} {\bibfield
  {journal} {\bibinfo  {journal} {Journal of Molecular Catalysis A: Chemical}\
  }\textbf {\bibinfo {volume} {138}},\ \bibinfo {pages} {211 } (\bibinfo {year}
  {1999})}\BibitemShut {NoStop}%
\bibitem[{\citenamefont{Zhang et~al.}(2009)\citenamefont{Zhang, Du, Petrik,
  Kimmel, Lyubinetsky, and Dohnalek}}]{zhang}
\bibinfo{author}{\bibfnamefont{Z.}~\bibnamefont{Zhang}},
  \bibinfo{author}{\bibfnamefont{Y.}~\bibnamefont{Du}},
  \bibinfo{author}{\bibfnamefont{N.~G.} \bibnamefont{Petrik}},
  \bibinfo{author}{\bibfnamefont{G.~A.} \bibnamefont{Kimmel}},
  \bibinfo{author}{\bibfnamefont{I.}~\bibnamefont{Lyubinetsky}},
  \bibnamefont{and}
  \bibinfo{author}{\bibfnamefont{Z.}~\bibnamefont{Dohnalek}},
  \bibinfo{journal}{The Journal of Physical Chemistry C}
  \textbf{\bibinfo{volume}{113}}, \bibinfo{pages}{1908} (\bibinfo{year}{2009})
  \BibitemShut {NoStop}%
\bibitem [{\citenamefont {Wang}\ \emph {et~al.}(2004)\citenamefont {Wang},
  \citenamefont {Gao}, \citenamefont {Kaltchev},\ and\ \citenamefont
  {Tysoe}}]{Wang2004}%
  \BibitemOpen
  \bibfield  {author} {\bibinfo {author} {\bibfnamefont {Y.}~\bibnamefont
  {Wang}}, \bibinfo {author} {\bibfnamefont {F.}~\bibnamefont {Gao}}, \bibinfo
  {author} {\bibfnamefont {M.}~\bibnamefont {Kaltchev}}, \ and\ \bibinfo
  {author} {\bibfnamefont {W.~T.}\ \bibnamefont {Tysoe}},\ }\href {\doibase
  10.1016/j.molcata.2003.08.009} {\bibfield  {journal} {\bibinfo  {journal}
  {Journal of Molecular Catalysis A: Chemical}\ }\textbf {\bibinfo {volume}
  {209}},\ \bibinfo {pages} {135 } (\bibinfo {year} {2004})}\BibitemShut
  {NoStop}%
\bibitem [{\citenamefont {Atodiresei}\ \emph {et~al.}(2009)\citenamefont
  {Atodiresei}, \citenamefont {Caciuc}, \citenamefont
  {Lazi\ifmmode~\acute{c}\else \'{c}\fi{}},\ and\ \citenamefont
  {Bl\"ugel}}]{Atodiresei2009}%
  \BibitemOpen
  \bibfield  {author} {\bibinfo {author} {\bibfnamefont {N.}~\bibnamefont
  {Atodiresei}}, \bibinfo {author} {\bibfnamefont {V.}~\bibnamefont {Caciuc}},
  \bibinfo {author} {\bibfnamefont {P.}~\bibnamefont
  {Lazi\ifmmode~\acute{c}\else \'{c}\fi{}}}, \ and\ \bibinfo {author}
  {\bibfnamefont {S.}~\bibnamefont {Bl\"ugel}},\ }\href {\doibase
  10.1103/PhysRevLett.102.136809} {\bibfield  {journal} {\bibinfo  {journal}
  {Phys. Rev. Lett.}\ }\textbf {\bibinfo {volume} {102}},\ \bibinfo {pages}
  {136809} (\bibinfo {year} {2009})}\BibitemShut {NoStop}%
\bibitem [{\citenamefont {Atodiresei}\ \emph {et~al.}(2010)\citenamefont
  {Atodiresei}, \citenamefont {Brede}, \citenamefont
  {Lazi\ifmmode~\acute{c}\else \'{c}\fi{}}, \citenamefont {Caciuc},
  \citenamefont {Hoffmann}, \citenamefont {R.},\ and\ \citenamefont
  {Bl\"ugel}}]{Atodiresei2010}%
  \BibitemOpen
  \bibfield  {author} {\bibinfo {author} {\bibfnamefont {N.}~\bibnamefont
  {Atodiresei}}, \bibinfo {author} {\bibfnamefont {J.}~\bibnamefont {Brede}},
  \bibinfo {author} {\bibfnamefont {P.}~\bibnamefont
  {Lazi\ifmmode~\acute{c}\else \'{c}\fi{}}}, \bibinfo {author} {\bibfnamefont
  {V.}~\bibnamefont {Caciuc}}, \bibinfo {author} {\bibfnamefont
  {G.}~\bibnamefont {Hoffmann}}, \bibinfo {author} {\bibfnamefont
  {R.}~\bibnamefont {Wiesendanger}}, \ and\ \bibinfo {author} {\bibfnamefont
  {S.}~\bibnamefont {Bl\"ugel}},\ }\href {\doibase
  10.1103/PhysRevLett.105.066601} {\bibfield  {journal} {\bibinfo  {journal}
  {Phys. Rev. Lett.}\ }\textbf {\bibinfo {volume} {105}},\ \bibinfo {pages}
  {066601} (\bibinfo {year} {2010})}\BibitemShut {NoStop}%
\bibitem [{\citenamefont {Bl\"ochl}(1994)}]{Bloechl1994}%
  \BibitemOpen
  \bibfield  {author} {\bibinfo {author} {\bibfnamefont {P.~E.}\ \bibnamefont
  {Bl\"ochl}},\ }\href@noop {} {\bibfield  {journal} {\bibinfo  {journal}
  {Phys. Rev. B}\ }\textbf {\bibinfo {volume} {50}},\ \bibinfo {pages} {17953}
  (\bibinfo {year} {1994})}\BibitemShut {NoStop}%
\bibitem [{\citenamefont {Kresse}\ and\ \citenamefont
  {Joubert}(1999)}]{Kresse1999}%
  \BibitemOpen
  \bibfield  {author} {\bibinfo {author} {\bibfnamefont {G.}~\bibnamefont
  {Kresse}}\ and\ \bibinfo {author} {\bibfnamefont {D.}~\bibnamefont
  {Joubert}},\ }\href {\doibase 10.1103/PhysRevB.59.1758} {\bibfield  {journal}
  {\bibinfo  {journal} {Phys. Rev. B}\ }\textbf {\bibinfo {volume} {59}},\
  \bibinfo {pages} {1758} (\bibinfo {year} {1999})}\BibitemShut {NoStop}%
\bibitem [{\citenamefont {Kresse}\ and\ \citenamefont
  {Furthm\"uller}(1996)}]{Kresse1996}%
  \BibitemOpen
  \bibfield  {author} {\bibinfo {author} {\bibfnamefont {G.}~\bibnamefont
  {Kresse}}\ and\ \bibinfo {author} {\bibfnamefont {J.}~\bibnamefont
  {Furthm\"uller}},\ }\href {\doibase 10.1016/0927-0256(96)00008-0} {\bibfield
  {journal} {\bibinfo  {journal} {Computational Materials Science}\ }\textbf
  {\bibinfo {volume} {6}},\ \bibinfo {pages} {15 } (\bibinfo {year}
  {1996})}\BibitemShut {NoStop}%
\bibitem [{\citenamefont {Kresse}\ and\ \citenamefont
  {Hafner}(1993)}]{Kresse1993}%
  \BibitemOpen
  \bibfield  {author} {\bibinfo {author} {\bibfnamefont {G.}~\bibnamefont
  {Kresse}}\ and\ \bibinfo {author} {\bibfnamefont {J.}~\bibnamefont
  {Hafner}},\ }\href {\doibase 10.1103/PhysRevB.47.558} {\bibfield  {journal}
  {\bibinfo  {journal} {Phys. Rev. B}\ }\textbf {\bibinfo {volume} {47}},\
  \bibinfo {pages} {558} (\bibinfo {year} {1993})}\BibitemShut {NoStop}%
\bibitem [{\citenamefont {Kresse}\ and\ \citenamefont
  {Hafner}(1994)}]{Kresse1994}%
  \BibitemOpen
  \bibfield  {author} {\bibinfo {author} {\bibfnamefont {G.}~\bibnamefont
  {Kresse}}\ and\ \bibinfo {author} {\bibfnamefont {J.}~\bibnamefont
  {Hafner}},\ }\href {\doibase 10.1103/PhysRevB.49.14251} {\bibfield  {journal}
  {\bibinfo  {journal} {Phys. Rev. B}\ }\textbf {\bibinfo {volume} {49}},\
  \bibinfo {pages} {14251} (\bibinfo {year} {1994})}\BibitemShut {NoStop}%
\bibitem [{\citenamefont {Perdew}\ \emph {et~al.}(1996)\citenamefont {Perdew},
  \citenamefont {Burke},\ and\ \citenamefont {Ernzerhof}}]{Perdew1996}%
  \BibitemOpen
  \bibfield  {author} {\bibinfo {author} {\bibfnamefont {J.~P.}\ \bibnamefont
  {Perdew}}, \bibinfo {author} {\bibfnamefont {K.}~\bibnamefont {Burke}}, \
  and\ \bibinfo {author} {\bibfnamefont {M.}~\bibnamefont {Ernzerhof}},\ }\href
  {\doibase 10.1103/PhysRevLett.77.3865} {\bibfield  {journal} {\bibinfo
  {journal} {Phys. Rev. Lett.}\ }\textbf {\bibinfo {volume} {77}},\ \bibinfo
  {pages} {3865} (\bibinfo {year} {1996})}\BibitemShut {NoStop}%
\bibitem [{\citenamefont {Perdew}\ \emph {et~al.}(1997)\citenamefont {Perdew},
  \citenamefont {Burke},\ and\ \citenamefont {Ernzerhof}}]{Perdew1997}%
  \BibitemOpen
  \bibfield  {author} {\bibinfo {author} {\bibfnamefont {J.~P.}\ \bibnamefont
  {Perdew}}, \bibinfo {author} {\bibfnamefont {K.}~\bibnamefont {Burke}}, \
  and\ \bibinfo {author} {\bibfnamefont {M.}~\bibnamefont {Ernzerhof}},\ }\href
  {\doibase 10.1103/PhysRevLett.78.1396} {\bibfield  {journal} {\bibinfo
  {journal} {Phys. Rev. Lett.}\ }\textbf {\bibinfo {volume} {78}},\ \bibinfo
  {pages} {1396} (\bibinfo {year} {1997})}\BibitemShut {NoStop}%
\bibitem [{\citenamefont {Grimme}(2006)}]{Grimme2006}%
  \BibitemOpen
  \bibfield  {author} {\bibinfo {author} {\bibfnamefont {S.}~\bibnamefont
  {Grimme}},\ }\href {\doibase 10.1002/jcc.20495} {\bibfield  {journal}
  {\bibinfo  {journal} {J. Comp. Chem.}\ }\textbf {\bibinfo {volume} {27}},\
  \bibinfo {pages} {1787} (\bibinfo {year} {2006})}\BibitemShut {NoStop}%
\bibitem [{\citenamefont {Wu}\ \emph {et~al.}(2001)\citenamefont {Wu},
  \citenamefont {Vargas}, \citenamefont {Nayak}, \citenamefont {Lotrich},\ and\
  \citenamefont {Scoles}}]{Wu2001}%
  \BibitemOpen
  \bibfield  {author} {\bibinfo {author} {\bibfnamefont {X.}~\bibnamefont
  {Wu}}, \bibinfo {author} {\bibfnamefont {M.~C.}\ \bibnamefont {Vargas}},
  \bibinfo {author} {\bibfnamefont {S.}~\bibnamefont {Nayak}}, \bibinfo
  {author} {\bibfnamefont {V.}~\bibnamefont {Lotrich}}, \ and\ \bibinfo
  {author} {\bibfnamefont {G.}~\bibnamefont {Scoles}},\ }\href {\doibase
  10.1063/1.1412004} {\bibfield  {journal} {\bibinfo  {journal} {J. Chem.
  Phys.}\ }\textbf {\bibinfo {volume} {115}},\ \bibinfo {pages} {8748}
  (\bibinfo {year} {2001})}\BibitemShut {NoStop}%
\bibitem [{\citenamefont {R.W.G.Wyckoff}(8 19)}]{Wyckoff1963}%
  \BibitemOpen
  \bibfield  {author} {\bibinfo {author} {\bibnamefont {R.W.G.Wyckoff}},\
  }\href@noop {} {\emph {\bibinfo {title} {Crystal Structures}}}\ (\bibinfo
  {publisher} {John Wiley \& Sons, New York, London},\ \bibinfo {year} {1963,
  Vol. 1,p. 318-19.})\BibitemShut {NoStop}%
\bibitem [{\citenamefont {Treutler}\ and\ \citenamefont
  {Ahlrichs}(1995)}]{Treutler1995}%
  \BibitemOpen
  \bibfield  {author} {\bibinfo {author} {\bibfnamefont {O.}~\bibnamefont
  {Treutler}}\ and\ \bibinfo {author} {\bibfnamefont {R.}~\bibnamefont
  {Ahlrichs}},\ }\href {\doibase 10.1063/1.469408} {\bibfield  {journal}
  {\bibinfo  {journal} {J. Chem. Phys.}\ }\textbf {\bibinfo {volume} {102}},\
  \bibinfo {pages} {346} (\bibinfo {year} {1995})}\BibitemShut {NoStop}%
\bibitem [{\citenamefont {Eichkorn}\ \emph {et~al.}(1997)\citenamefont
  {Eichkorn}, \citenamefont {Weigend}, \citenamefont {Treutler},\ and\
  \citenamefont {Ahlrichs}}]{Eichkorn1997}%
  \BibitemOpen
  \bibfield  {author} {\bibinfo {author} {\bibfnamefont {K.}~\bibnamefont
  {Eichkorn}}, \bibinfo {author} {\bibfnamefont {F.}~\bibnamefont {Weigend}},
  \bibinfo {author} {\bibfnamefont {O.}~\bibnamefont {Treutler}}, \ and\
  \bibinfo {author} {\bibfnamefont {R.}~\bibnamefont {Ahlrichs}},\ }\href
  {\doibase 10.1007/s002140050244} {\bibfield  {journal} {\bibinfo  {journal}
  {Theor. Chem. Acc.}\ }\textbf {\bibinfo {volume} {97}},\ \bibinfo {pages}
  {119} (\bibinfo {year} {1997})}\BibitemShut {NoStop}%
\bibitem [{\citenamefont {Jiang}\ and\ \citenamefont
  {Carter}(2005)}]{Jiang2005}%
  \BibitemOpen
  \bibfield  {author} {\bibinfo {author} {\bibfnamefont {D.~E.}\ \bibnamefont
  {Jiang}}\ and\ \bibinfo {author} {\bibfnamefont {E.~A.}\ \bibnamefont
  {Carter}},\ }\href {\doibase 10.1103/PhysRevB.72.165410} {\bibfield
  {journal} {\bibinfo  {journal} {Phys. Rev. B}\ }\textbf {\bibinfo {volume}
  {72}},\ \bibinfo {pages} {165410} (\bibinfo {year} {2005})}\BibitemShut
  {NoStop}%
\bibitem [{\citenamefont {Wehling}\ \emph {et~al.}(2008)\citenamefont
  {Wehling}, \citenamefont {Lichtenstein},\ and\ \citenamefont
  {Katsnelson}}]{Wehling2008}%
  \BibitemOpen
  \bibfield  {author} {\bibinfo {author} {\bibfnamefont {T.~O.}\ \bibnamefont
  {Wehling}}, \bibinfo {author} {\bibfnamefont {A.~I.}\ \bibnamefont
  {Lichtenstein}}, \ and\ \bibinfo {author} {\bibfnamefont {M.~I.}\
  \bibnamefont {Katsnelson}},\ }\href {\doibase 10.1063/1.3033202} {\bibfield
  {journal} {\bibinfo  {journal} {Appl. Phys. Lett.}\ }\textbf {\bibinfo
  {volume} {93}},\ \bibinfo {eid} {202110} (\bibinfo {year}
  {2008})}\BibitemShut {NoStop}%
\bibitem [{\citenamefont {Vign{\'e}-Maeder}\ and\ \citenamefont
  {Sautet}(1997)}]{VigneMaeder1997}%
  \BibitemOpen
  \bibfield  {author} {\bibinfo {author} {\bibfnamefont {F.}~\bibnamefont
  {Vign{\'e}-Maeder}}\ and\ \bibinfo {author} {\bibfnamefont {P.}~\bibnamefont
  {Sautet}},\ }\href {\doibase 10.1021/jp964012b} {\bibfield  {journal}
  {\bibinfo  {journal} {J. Phys. Chem. B}\ }\textbf {\bibinfo {volume} {101}},\
  \bibinfo {pages} {8197} (\bibinfo {year} {1997})}\BibitemShut {NoStop}%
\bibitem [{\citenamefont {Ceresoli}\ \emph {et~al.}(2000)\citenamefont
  {Ceresoli}, \citenamefont {Bernasconi}, \citenamefont {Iarlori},
  \citenamefont {Parrinello},\ and\ \citenamefont {Tosatti}}]{Ceresoli2000}%
  \BibitemOpen
  \bibfield  {author} {\bibinfo {author} {\bibfnamefont {D.}~\bibnamefont
  {Ceresoli}}, \bibinfo {author} {\bibfnamefont {M.}~\bibnamefont
  {Bernasconi}}, \bibinfo {author} {\bibfnamefont {S.}~\bibnamefont {Iarlori}},
  \bibinfo {author} {\bibfnamefont {M.}~\bibnamefont {Parrinello}}, \ and\
  \bibinfo {author} {\bibfnamefont {E.}~\bibnamefont {Tosatti}},\ }\href
  {\doibase 10.1103/PhysRevLett.84.3887} {\bibfield  {journal} {\bibinfo
  {journal} {Phys. Rev. Lett.}\ }\textbf {\bibinfo {volume} {84}},\ \bibinfo
  {pages} {3887} (\bibinfo {year} {2000})}\BibitemShut {NoStop}%
\bibitem [{\citenamefont {Ricci}\ and\ \citenamefont
  {Pacchioni}(2004)}]{Ricci2004}%
  \BibitemOpen
  \bibfield  {author} {\bibinfo {author} {\bibfnamefont {D.}~\bibnamefont
  {Ricci}}\ and\ \bibinfo {author} {\bibfnamefont {G.}~\bibnamefont
  {Pacchioni}},\ }\href {\doibase 10.1103/PhysRevB.69.161307} {\bibfield
  {journal} {\bibinfo  {journal} {Phys. Rev. B}\ }\textbf {\bibinfo {volume}
  {69}},\ \bibinfo {pages} {161307(R)} (\bibinfo {year} {2004})}\BibitemShut
  {NoStop}%
\bibitem [{\citenamefont {Peri}\ and\ \citenamefont
  {Hensley~Jr.}(1968)}]{Peri1968}%
  \BibitemOpen
  \bibfield  {author} {\bibinfo {author} {\bibfnamefont {J.~B.}\ \bibnamefont
  {Peri}}\ and\ \bibinfo {author} {\bibfnamefont {A.~L.}\ \bibnamefont
  {Hensley~Jr.}},\ }\href {\doibase 10.1021/j100854a041} {\bibfield  {journal}
  {\bibinfo  {journal} {J. Phys. Chem.}\ }\textbf {\bibinfo {volume} {72}},\
  \bibinfo {pages} {2926} (\bibinfo {year} {1968})}\BibitemShut {NoStop}%
\bibitem [{\citenamefont {Sindorf}\ and\ \citenamefont
  {Maciel}(1983)}]{Sindorf1983}%
  \BibitemOpen
  \bibfield  {author} {\bibinfo {author} {\bibfnamefont {D.~W.}\ \bibnamefont
  {Sindorf}}\ and\ \bibinfo {author} {\bibfnamefont {G.~E.}\ \bibnamefont
  {Maciel}},\ }\href {\doibase 10.1021/ja00344a012} {\bibfield  {journal}
  {\bibinfo  {journal} {J. Am. Chem. Soc.}\ }\textbf {\bibinfo {volume}
  {105}},\ \bibinfo {pages} {1487} (\bibinfo {year} {1983})}\BibitemShut
  {NoStop}%
\bibitem [{\citenamefont {Puhakka}\ \emph {et~al.}(2011)\citenamefont
  {Puhakka}, \citenamefont {M.Riihim\"aki}, \citenamefont {P\"a\"akk\"onen},\
  and\ \citenamefont {Keiski}}]{Puhakka2011}%
  \BibitemOpen
  \bibfield  {author} {\bibinfo {author} {\bibfnamefont {E.}~\bibnamefont
  {Puhakka}}, \bibinfo {author} {\bibnamefont {M.Riihim\"aki}}, \bibinfo
  {author} {\bibfnamefont {M.~T.}\ \bibnamefont {P\"a\"akk\"onen}}, \ and\
  \bibinfo {author} {\bibfnamefont {R.~L.}\ \bibnamefont {Keiski}},\ }\href
  {\doibase 10.1080/01457632.2010.495626} {\bibfield  {journal} {\bibinfo
  {journal} {Heat Transfer Engineering}\ }\textbf {\bibinfo {volume} {32}},\
  \bibinfo {pages} {282} (\bibinfo {year} {2011})}\BibitemShut {NoStop}%
\bibitem [{\citenamefont {Wright}\ and\ \citenamefont
  {Leadbetter}(1975)}]{Wright1975}%
  \BibitemOpen
  \bibfield  {author} {\bibinfo {author} {\bibfnamefont {A.~F.}\ \bibnamefont
  {Wright}}\ and\ \bibinfo {author} {\bibfnamefont {A.~J.}\ \bibnamefont
  {Leadbetter}},\ }\href {\doibase 10.1080/00318087508228690} {\bibfield
  {journal} {\bibinfo  {journal} {Philos. Mag.}\ }\textbf {\bibinfo {volume}
  {31}},\ \bibinfo {pages} {1391} (\bibinfo {year} {1975})}\BibitemShut
  {NoStop}%
\bibitem [{\citenamefont {Yang}\ \emph {et~al.}(2005)\citenamefont {Yang},
  \citenamefont {Meng}, \citenamefont {Xu},\ and\ \citenamefont
  {Wang}}]{Yang2005}%
  \BibitemOpen
  \bibfield  {author} {\bibinfo {author} {\bibfnamefont {J.}~\bibnamefont
  {Yang}}, \bibinfo {author} {\bibfnamefont {S.}~\bibnamefont {Meng}}, \bibinfo
  {author} {\bibfnamefont {L.}~\bibnamefont {Xu}}, \ and\ \bibinfo {author}
  {\bibfnamefont {E.~G.}\ \bibnamefont {Wang}},\ }\href {\doibase
  10.1103/PhysRevB.71.035413} {\bibfield  {journal} {\bibinfo  {journal} {Phys.
  Rev. B}\ }\textbf {\bibinfo {volume} {71}},\ \bibinfo {pages} {035413}
  (\bibinfo {year} {2005})}\BibitemShut {NoStop}%
\bibitem [{\citenamefont {Rosa}\ \emph {et~al.}(1999)\citenamefont {Rosa},
  \citenamefont {Baerends}, \citenamefont {van Gisbergen}, \citenamefont {van
  Lenthe}, \citenamefont {Groeneveld},\ and\ \citenamefont
  {Snijders}}]{Rosa1999}%
  \BibitemOpen
  \bibfield  {author} {\bibinfo {author} {\bibfnamefont {A.}~\bibnamefont
  {Rosa}}, \bibinfo {author} {\bibfnamefont {E.~J.}\ \bibnamefont {Baerends}},
  \bibinfo {author} {\bibfnamefont {S.~J.~A.}\ \bibnamefont {van Gisbergen}},
  \bibinfo {author} {\bibfnamefont {E.}~\bibnamefont {van Lenthe}}, \bibinfo
  {author} {\bibfnamefont {J.~A.}\ \bibnamefont {Groeneveld}}, \ and\ \bibinfo
  {author} {\bibfnamefont {J.~G.}\ \bibnamefont {Snijders}},\ }\href {\doibase
  10.1021/ja990747t} {\bibfield  {journal} {\bibinfo  {journal} {J. Am. Chem.
  Soc.}\ }\textbf {\bibinfo {volume} {121}},\ \bibinfo {pages} {10356}
  (\bibinfo {year} {1999})}\BibitemShut {NoStop}%
\bibitem [{\citenamefont {Liu}\ \emph {et~al.}(2010)\citenamefont {Liu},
  \citenamefont {Ning}, \citenamefont {Luo}, \citenamefont {Shi},\ and\
  \citenamefont {Deng}}]{Liu2010}%
  \BibitemOpen
  \bibfield  {author} {\bibinfo {author} {\bibfnamefont {K.}~\bibnamefont
  {Liu}}, \bibinfo {author} {\bibfnamefont {C.~G.}\ \bibnamefont {Ning}},
  \bibinfo {author} {\bibfnamefont {Z.~H.}\ \bibnamefont {Luo}}, \bibinfo
  {author} {\bibfnamefont {L.~L.}\ \bibnamefont {Shi}}, \ and\ \bibinfo
  {author} {\bibfnamefont {J.~K.}\ \bibnamefont {Deng}},\ }\href {\doibase
  10.1016/j.cplett.2010.08.003} {\bibfield  {journal} {\bibinfo  {journal}
  {Chem. Phys. Lett.}\ }\textbf {\bibinfo {volume} {497}},\ \bibinfo {pages}
  {229 } (\bibinfo {year} {2010})}\BibitemShut {NoStop}%
\bibitem [{\citenamefont {Brockway}\ \emph {et~al.}(1938)\citenamefont
  {Brockway}, \citenamefont {Ewens},\ and\ \citenamefont
  {Lister}}]{Brockway1938}%
  \BibitemOpen
  \bibfield  {author} {\bibinfo {author} {\bibfnamefont {L.~O.}\ \bibnamefont
  {Brockway}}, \bibinfo {author} {\bibfnamefont {R.~V.~G.}\ \bibnamefont
  {Ewens}}, \ and\ \bibinfo {author} {\bibfnamefont {M.~W.}\ \bibnamefont
  {Lister}},\ }\href {\doibase 10.1039/TF9383401350} {\bibfield  {journal}
  {\bibinfo  {journal} {Trans. Faraday Soc.}\ }\textbf {\bibinfo {volume}
  {34}},\ \bibinfo {pages} {1350} (\bibinfo {year} {1938})}\BibitemShut
  {NoStop}%
\bibitem [{\citenamefont {Li}\ \emph {et~al.}(1995)\citenamefont {Li},
  \citenamefont {Schreckenbach},\ and\ \citenamefont {Ziegler}}]{Li1995}%
  \BibitemOpen
  \bibfield  {author} {\bibinfo {author} {\bibfnamefont {J.}~\bibnamefont
  {Li}}, \bibinfo {author} {\bibfnamefont {G.}~\bibnamefont {Schreckenbach}}, \
  and\ \bibinfo {author} {\bibfnamefont {T.}~\bibnamefont {Ziegler}},\ }\href
  {\doibase 10.1021/ja00106a056} {\bibfield  {journal} {\bibinfo  {journal} {J.
  Am. Chem. Soc.}\ }\textbf {\bibinfo {volume} {117}},\ \bibinfo {pages} {486}
  (\bibinfo {year} {1995})}\BibitemShut {NoStop}%
\bibitem [{\citenamefont {Myllyoja}\ \emph {et~al.}(1999)\citenamefont
  {Myllyoja}, \citenamefont {Suvanto}, \citenamefont {Kurhinen}, \citenamefont
  {Hirva},\ and\ \citenamefont {Pakkanen}}]{Myllyoja1999}%
  \BibitemOpen
  \bibfield  {author} {\bibinfo {author} {\bibfnamefont {S.}~\bibnamefont
  {Myllyoja}}, \bibinfo {author} {\bibfnamefont {M.}~\bibnamefont {Suvanto}},
  \bibinfo {author} {\bibfnamefont {M.}~\bibnamefont {Kurhinen}}, \bibinfo
  {author} {\bibfnamefont {P.}~\bibnamefont {Hirva}}, \ and\ \bibinfo {author}
  {\bibfnamefont {T.~A.}\ \bibnamefont {Pakkanen}},\ }\href {\doibase
  10.1016/S0039-6028(99)00882-1} {\bibfield  {journal} {\bibinfo  {journal}
  {Surf. Sci.}\ }\textbf {\bibinfo {volume} {441}},\ \bibinfo {pages} {454 }
  (\bibinfo {year} {1999})}\BibitemShut {NoStop}%
\bibitem [{\citenamefont {Baur}(1978)}]{Baur1978}%
  \BibitemOpen
  \bibfield  {author} {\bibinfo {author} {\bibfnamefont {W.~H.}\ \bibnamefont
  {Baur}},\ }\href {\doibase 10.1107/S0567740878006640} {\bibfield  {journal}
  {\bibinfo  {journal} {Acta Crystallogr. Sect. B}\ }\textbf {\bibinfo {volume}
  {34}},\ \bibinfo {pages} {1751} (\bibinfo {year} {1978})}\BibitemShut
  {NoStop}%
\bibitem[{\citenamefont{Morrow and Cody}(1976)}]{morrow}
\bibinfo{author}{\bibfnamefont{B.~A.} \bibnamefont{Morrow}} \bibnamefont{and}
  \bibinfo{author}{\bibfnamefont{I.~A.} \bibnamefont{Cody}},
  \bibinfo{journal}{The Journal of Physical Chemistry}
  \textbf{\bibinfo{volume}{80}}, \bibinfo{pages}{1995} (\bibinfo{year}{1976})
  \BibitemShut 
  {NoStop}%
\bibitem[{\citenamefont{Zecchina and Aréan}(1993)}]{adriano}
\bibinfo{author}{\bibfnamefont{A.}~\bibnamefont{Zecchina}} \bibnamefont{and}
  \bibinfo{author}{\bibfnamefont{C.~O.} \bibnamefont{Aréan}},
  \bibinfo{journal}{Catalysis Reviews} \textbf{\bibinfo{volume}{35}},
  \bibinfo{pages}{261} (\bibinfo{year}{1993})
  \BibitemShut 
  {NoStop}%
\bibitem [{\citenamefont {Domenichini}\ \emph {et~al.}(2008)\citenamefont
  {Domenichini}, \citenamefont {Prunier}, \citenamefont {Petukov},
  \citenamefont {Li}, \citenamefont {M{\o}ller},\ and\ \citenamefont
  {Bourgeois}}]{Domenichini2008}%
  \BibitemOpen
  \bibfield  {author} {\bibinfo {author} {\bibfnamefont {B.}~\bibnamefont
  {Domenichini}}, \bibinfo {author} {\bibfnamefont {J.}~\bibnamefont
  {Prunier}}, \bibinfo {author} {\bibfnamefont {M.}~\bibnamefont {Petukov}},
  \bibinfo {author} {\bibfnamefont {Z.}~\bibnamefont {Li}}, \bibinfo {author}
  {\bibfnamefont {P.~J.}\ \bibnamefont {M{\o}ller}}, \ and\ \bibinfo {author}
  {\bibfnamefont {S.}~\bibnamefont {Bourgeois}},\ }\href {\doibase
  10.1016/j.elspec.2008.02.002} {\bibfield  {journal} {\bibinfo  {journal} {J.
  Electron Spectrosc. Relat. Phenom.}\ }\textbf {\bibinfo {volume} {163}},\
  \bibinfo {pages} {19 } (\bibinfo {year} {2008})}\BibitemShut {NoStop}%
\bibitem [{\citenamefont {Lewis}\ \emph {et~al.}(1984)\citenamefont {Lewis},
  \citenamefont {Golden},\ and\ \citenamefont {Smith}}]{Lewis1984}%
  \BibitemOpen
  \bibfield  {author} {\bibinfo {author} {\bibfnamefont {K.~E.}\ \bibnamefont
  {Lewis}}, \bibinfo {author} {\bibfnamefont {D.~M.}\ \bibnamefont {Golden}}, \
  and\ \bibinfo {author} {\bibfnamefont {G.~P.}\ \bibnamefont {Smith}},\ }\href
  {\doibase 10.1021/ja00326a004} {\bibfield  {journal} {\bibinfo  {journal} {J.
  Am. Chem. Soc.}\ }\textbf {\bibinfo {volume} {106}},\ \bibinfo {pages} {3905}
  (\bibinfo {year} {1984})}\BibitemShut {NoStop}%
\bibitem [{\citenamefont {Sniatynsky}\ and\ \citenamefont
  {Cede\={n}o}(2004)}]{Sniatynsky2004}%
  \BibitemOpen
  \bibfield  {author} {\bibinfo {author} {\bibfnamefont {R.}~\bibnamefont
  {Sniatynsky}}\ and\ \bibinfo {author} {\bibfnamefont {D.~L.}\ \bibnamefont
  {Cede\={n}o}},\ }\href {\doibase 10.1016/j.theochem.2004.09.018} {\bibfield
  {journal} {\bibinfo  {journal} {Journal of Molecular Structure: THEOCHEM}\
  }\textbf {\bibinfo {volume} {711}},\ \bibinfo {pages} {123 } (\bibinfo {year}
  {2004})}\BibitemShut {NoStop}%
\bibitem [{\citenamefont {Kim}\ \emph {et~al.}(2003)\citenamefont {Kim},
  \citenamefont {Ha},\ and\ \citenamefont {Kim}}]{Kiminami2003}%
  \BibitemOpen
  \bibfield  {author} {\bibinfo {author} {\bibfnamefont {J.~C.}\ \bibnamefont
  {Kim}}, \bibinfo {author} {\bibfnamefont {G.~H.}\ \bibnamefont {Ha}}, \ and\
  \bibinfo {author} {\bibfnamefont {B.~K.}\ \bibnamefont {Kim}},\ }\href
  {\doibase 10.4028/www.scientific.net/JMNM.20-21.237} {\bibfield  {journal}
  {\bibinfo  {journal} {J. Metastable and Nanocrystalline Mater.}\ }\textbf
  {\bibinfo {volume} {20-21}},\ \bibinfo {pages} {237} (\bibinfo {year}
  {2003})}\BibitemShut {NoStop}%
\bibitem [{\citenamefont {Huth}()}]{HuthPrivate}%
  \BibitemOpen
  \bibfield  {author} {\bibinfo {author} {\bibfnamefont {M.}~\bibnamefont
  {Huth}},\ }\href@noop {} {}\bibinfo {note} {private
  communication}\BibitemShut {NoStop}%
\end{thebibliography}

%
\end{document}